\documentclass[aps,amsmath,amssymb,reprint,longbibliography]{revtex4-2}
\usepackage{graphicx}
\usepackage{dcolumn}
\usepackage{bm,upgreek}
\usepackage{subcaption}

\usepackage[utf8]{inputenc}
\usepackage[T1]{fontenc}
\usepackage{mathptmx}
\usepackage{ragged2e}
\usepackage{siunitx}
\usepackage{float}

\begin{document}

\preprint{APS/123-QED}

\title{Photonic crystal cavities based on suspended yttrium iron garnet nanobeams}

\author{A. Rashedi}

\author{M. Ebrahimi}

\author{Y. Huang}
\affiliation{Department of Physics, University of Alberta, Edmonton, Alberta T6G 2E9, Canada}

\author{V. A. S. V. Bittencourt}
\affiliation{ISIS (UMR 7006), Universit\'{e} de Strasbourg, 67000 Strasbourg, France}

\author{M. J. Rudd}

\author{J.P. Davis}
\email{jdavis@ualberta.ca}
\affiliation{Department of Physics, University of Alberta, Edmonton, Alberta T6G 2E9, Canada}


\begin{abstract}
We report the fabrication and optical characterization of an air-suspended photonic crystal nanobeam cavity in yttrium-iron-garnet (YIG) realized by focused-ion-beam milling. YIG's combination of low optical loss and ferrimagnetism makes it highly attractive for quantum technologies, yet prior work has largely been focused on millimeter-scale spheres and simple microstructures, hindering true on-chip integration. Demonstrating nanometer-scale patterning in a suspended geometry therefore represents an important advance. Finite-element simulations predict that the same structure supports a flapping-type mechanical mode at $\Omega_{\,m} / 2\pi \approx \qty{1.52}{GHz}$ and a backward-volume spin-wave mode at $\omega_{\text{mag}}/2\pi = \qty{11.59}{GHz}$ under an in-plane bias field. Although we measure only the photonic resonance (intrinsic Q $\approx 2 \times 10^3$) in this study, the device lays the groundwork for future exploration of coupled photon–phonon–magnon dynamics once higher optical quality factors are achieved.

\end{abstract}

\maketitle

Hybrid quantum systems are emerging as a promising pathway in the advancement of quantum technologies. By integrating the strengths of individual systems while mitigating their weaknesses, these platforms offer enhanced capabilities and functionalities \cite{Lachance-Quirion2019}. Optomechanical and magnomechanical platforms are prime examples of such systems. Optomechanical systems \cite{Aspelmeyer2014, Gröblacher2009, Barzanjeh2022}, in particular, combine optical and mechanical resonators to perform a variety of tasks, ranging from dark matter detection \cite{Abbott2022, Manley2020, Baker2024, Vadakkumbatt2021, Hirschel2024} to ultra-sensitive torque and magnetic measurements \cite{Kim2017, Rudd2019, Losby2018}, as well as coherent transduction of microwave photons to optical photons \cite{Jiang2020, Forsch2020, Ramp2020, Arnold2020, Weaver2024, Brubaker2022} and numerous other applications \cite{Safavi-Naeini2013, Riedinger2018, Riedinger2016, Hong2017, Wallucks2020}. These systems can offer similar functionality, but with a  smaller footprint, when compared to other platforms \cite{Smith2023, Tu2022}.

An important class within optomechanical systems is the optomechanical crystal (OMC) \cite{Eichenfield2009, Safavi-Naeini2014, Ren2020, Benevides2017, Gomis-Bresco2014}. These structures consist of photonic and phononic crystal cavities that simultaneously confine optical and mechanical modes within a small volume. Such nanostructures are incredibly versatile and can be precisely engineered for specific applications, enhancing the interaction between light and mechanical vibrations. This coupling is useful for quantum information processing, precision sensing, and other quantum technologies. Optomechanical crystals have been realized in various materials, including silicon, silicon nitride, gallium arsenide, and diamond \cite{Eichenfield2009, Amir2011, Grutter2015, Ramp2020, Burek2016}.

Magnomechanical systems, another type of hybrid system, focus on interactions between magnons, the quanta of spin waves, and phonons \cite{Zuo2024}. Yttrium iron garnet (YIG) is the preferred material for magnonic systems \cite{Serga2010} due to its low Gilbert damping \cite{Klingler2017} ($\alpha \approx 10^{-5}$) and low optical absorption in the infrared region \cite{Wemple1974}. YIG macroscopic spheres have been extensively studied \cite{Potts2020, Potts2023, Bittencourt2023, Liu2023, Li2021, Zhang2016, Zhang2014, Zhang2015}. However, the limited mode overlap between phononic and magnonic modes in YIG spheres restricts the single phonon–magnon coupling rate to the millihertz range, rendering it too weak for practical quantum technology applications. To overcome this limitation, nano- and microfabrication techniques are essential to enhance the mode overlap and, consequently, increase the interaction rate. Realizing an optomechanical crystal in a magnetic material like YIG promises to bridge the gap between optomechanical and magnomechanical systems. This integration would enable the creation of a multiparticle system capable of addressing various contemporary challenges, such as achieving near-unity efficiency in the coherent conversion of microwave photons to telecommunication photons \cite{Engelhardt2022, Simon2017}.

One primary challenge to such a platform is that YIG is not compatible with the conventional fabrication methods used for other materials hosting optomechanical crystals, such as silicon and silicon nitride. Despite significant efforts toward fabricating micro- and nanostructures in YIG, these attempts have generally resulted in simple geometries that aim to support only a single type of mode, either optical or mechanical, rather than simultaneous support for both \cite {Zhu2022, Trempler2020, Schmidt2020, Arisawa2019, Zhu2017, Heyroth2019,Taniguchi2024}. Magnetophotonic crystals \cite{Inoue2006, Fakhrul2019, Inoue2013} in YIG have been pursued for decades, but they generally use planar multilayer stacks aimed at static magneto-optical effects and lack the suspended geometries required to confine mechanical modes and enable dynamic interactions among the three quasiparticles \cite{Chakravarty2011_YIGPHC}.

In this work, we report the first observation of optical resonance in a nanofabricated suspended structure made from YIG, designed to support three types of bosonic quasiparticles: optical photons, phonons, and magnons. By confining all three within a shared volume and maximizing their mode overlap, the device has the potential to enable mutual interactions. This lays the groundwork for functional magneto-optomechanical systems.

\begin{figure}[t]
\includegraphics[width= 8.5 cm]{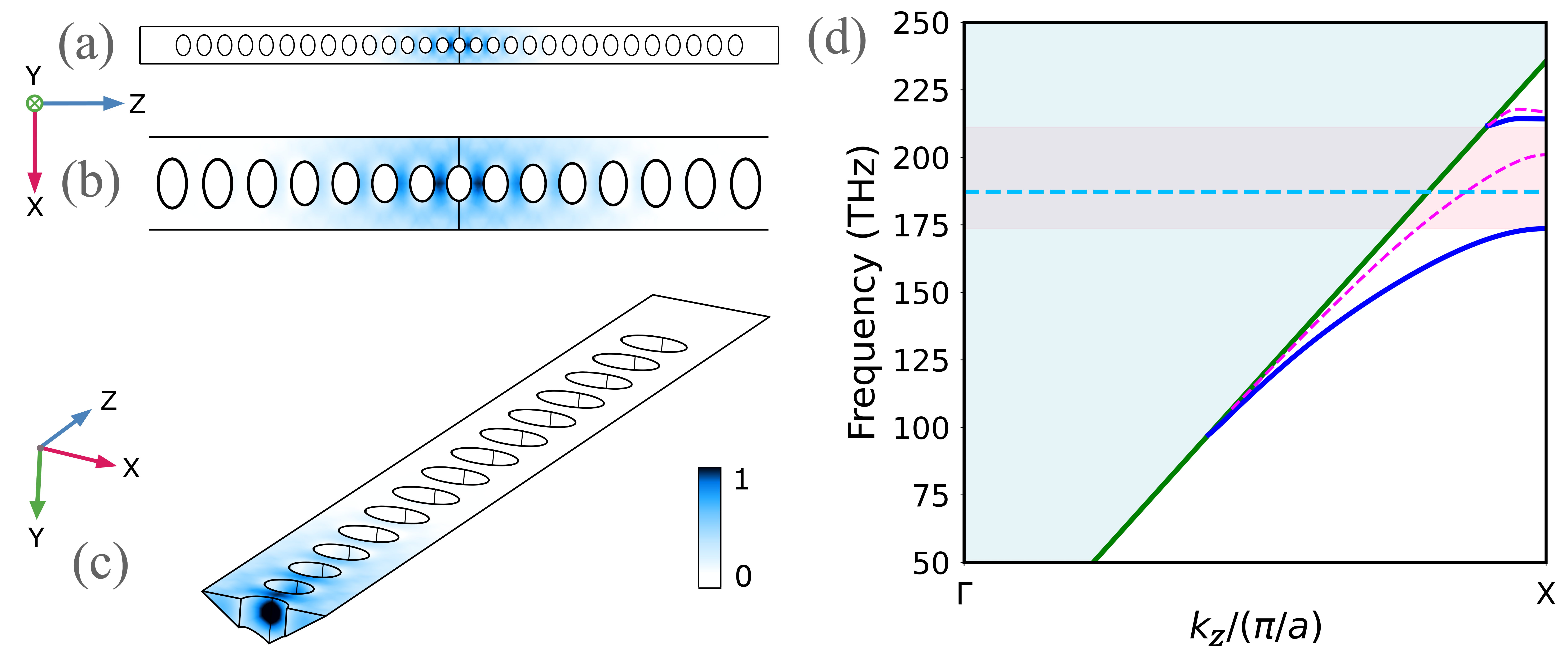}
\captionsetup{justification=justified}
\caption{\justifying \small Design of the YIG optomechanical crystal (OMC). (a)-(c) Amplitude profile of the confined optical mode's electric field, illustrating that most of the field is concentrated in the center of the beam and within the YIG, which is essential for achieving a high phonon-photon-magnon coupling rate. (b) A zoomed view of the cavity area of the waveguide. (c) A cross-sectional view of the center of the waveguide. (d) Optical band diagram of the OMC, where the solid blue lines represent the guided transverse electric (TE-like) modes, and the dashed magenta lines depict the guided transverse magnetic (TM-like) modes of the waveguide. The shaded pink region indicates the optical semi-band gap, intersected by the light line (green solid line), while the continuous leaky modes coupling to air are represented by the shaded blue area above the light line. The confined mode at the edge of the first Brillouin zone is highlighted by the cyan horizontal dashed line.}
\label{Fig:01}
\end{figure}

\section{Design and Fabrication}

\begin{figure}[b]
\includegraphics[width= 8.3 cm]{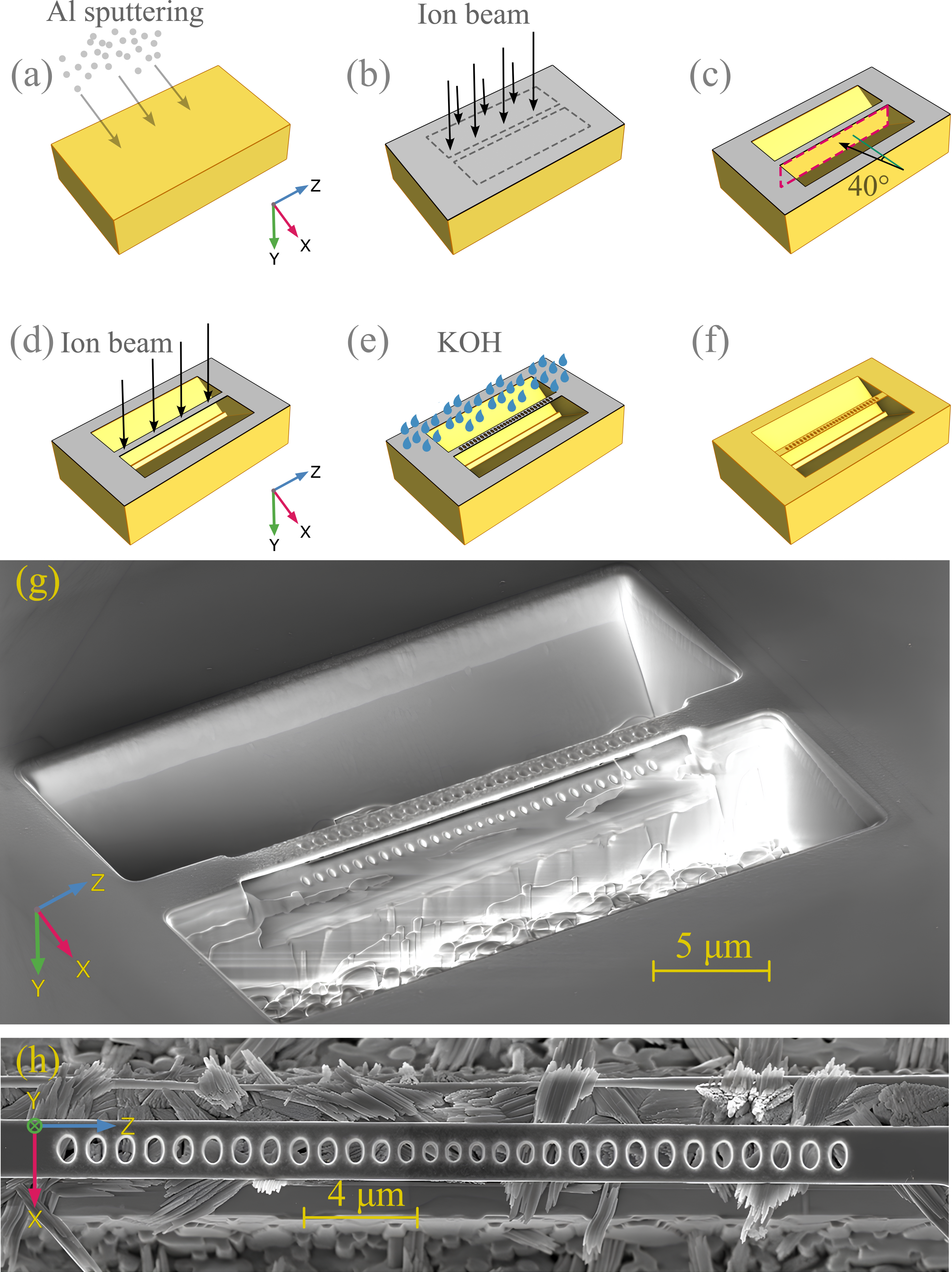}
\captionsetup{justification=justified}
\caption{\justifying \small (a)-(f) Fabrication process flow of the YIG optomechanical crystal cavity (OMC). (a) The process begins with the deposition of a sacrificial aluminum (Al) layer on the YIG substrate. (b)-(d) The Xe ion beam is then used for milling the trenches in a three-step process: initial coarse trench milling at 1–3 nA and 30 kV, followed by intermediate shaping at 100–300 pA, and finally a polishing step at approximately 30 pA. This sequence minimizes sidewall roughness, precisely shapes the beam, and accurately drills the elliptical holes. (e) The final step involves removing the Al layer using a potassium hydroxide (KOH) wet etch to reveal the completed YIG OMC. (f) The result is a suspended structure ready for characterization. (g) Scanning electron micrograph (SEM) of the prepared device before the Al removal step. (h) SEM of the finished device.}
\label{Fig:02}
\end{figure}

To realize a platform that hosts optical, mechanical, and magnon modes simultaneously, we began by designing an OMC similar to the diamond OMC of Burek et al.~\cite{Burek2016} and Babinec et al.~\cite{Babinec2011}. The device consists of a suspended rectangular YIG nanobeam patterned along its length by elliptical holes of varying dimensions, forming a photonic crystal waveguide. It is doubly clamped by the YIG substrate at both ends, as shown in Fig.~\ref{Fig:01}(a)-(c) and Fig.~\ref{Fig:02}(f). We optimized the optical bandgap of the device via finite element method (FEM) simulations in COMSOL Multiphysics \cite{COMSOL} to confine only a single optical mode at \(\omega_{\,0} / 2\pi \approx \qty{187}{THz} \, (\sim \qty{1600}{nm})\), depicted by the cyan horizontal dashed line deep within the optical bandgap, as shown in Fig.~\ref{Fig:01}(d). All higher-order defect cell modes lie outside this gap, ensuring that only the desired optical mode is confined.

Modes with most of their electric field perpendicular to the beam axis (TE-like) are shown with solid blue lines, and modes with most of their electric field parallel to the beam axis (TM-like) are shown with dashed magenta lines in Fig.~\ref{Fig:01}(d). Due to symmetry and polarization mismatch, TE-like and TM-like modes do not couple to one another, enabling a large bandgap of \qty{40}{THz} for TE-like modes. The electric field amplitude profile of the confined optical mode at the edge of the first Brillouin zone is shown in Fig.~\ref{Fig:01}(a)-(c).

As previously mentioned, YIG does not yet have standardized processes compatible with conventional microfabrication techniques. Therefore, we turn to focused ion beam (FIB) milling, which uses high-energy ion beams to physically remove material, sculpting the final design from the substrate, effectively functioning as a nanoscale computer numerical control (CNC) machine \cite{Kim2012}. FIB has also been used to realize high-performance photonic-crystal nanobeam cavities in other platforms; for example, in lithium-niobate-on-insulator a loaded quality factor of \(7.3\times 10^{4}\) has been measured~\cite{lee2024highqv}, which supports the suitability of this approach for our geometry.
 The high-energy nature of the ion beam introduces several challenges that can degrade the quality of the final structure. During the milling process, issues such as heat generation, ion implantation, and re-deposition of sputtered material can significantly affect the precision and functionality of the final device.

The localized heating from the ion beam can alter the material properties of YIG or cause structural deformation, impacting the device's functionality. Additionally, ion implantation, where some ions from the beam embed into the YIG substrate, can modify its magnetic or optical properties, potentially introducing defects \cite{Khizroev2004, Fraser2010, Losby2015}. Furthermore, material redeposition, in which removed material settles onto unwanted areas, can lead to excess surface roughness and otherwise contaminate high-precision regions. 

Our substrate consists of an \qty{840}{nm} thick layer of YIG with a (111) crystallographic orientation on a \qty{500}{\micro \metre} thick gadolinium gallium garnet (GGG) base. To mitigate the issues associated with FIB milling, we introduce a sacrificial aluminum layer deposited on the YIG surface at the outset. Aluminum was chosen for several reasons. First, it is commonly used as a hard mask in Si and SiN fabrication and can be easily removed using standard wet etching processes. Second, aluminum effectively conducts heat, aiding in heat dissipation and thereby reducing the risk of heat-induced damage to the YIG substrate. Third, during the milling process, any redeposited material settles on this sacrificial layer, which can then be easily removed along with the Al layer, ensuring a cleaner final structure. Finally, this aluminum layer serves as a barrier, reducing the implantation of FIB ions into the substrate and preserving the integrity of the YIG layer. 

\begin{figure}[t]
\includegraphics[width= 7.5 cm]{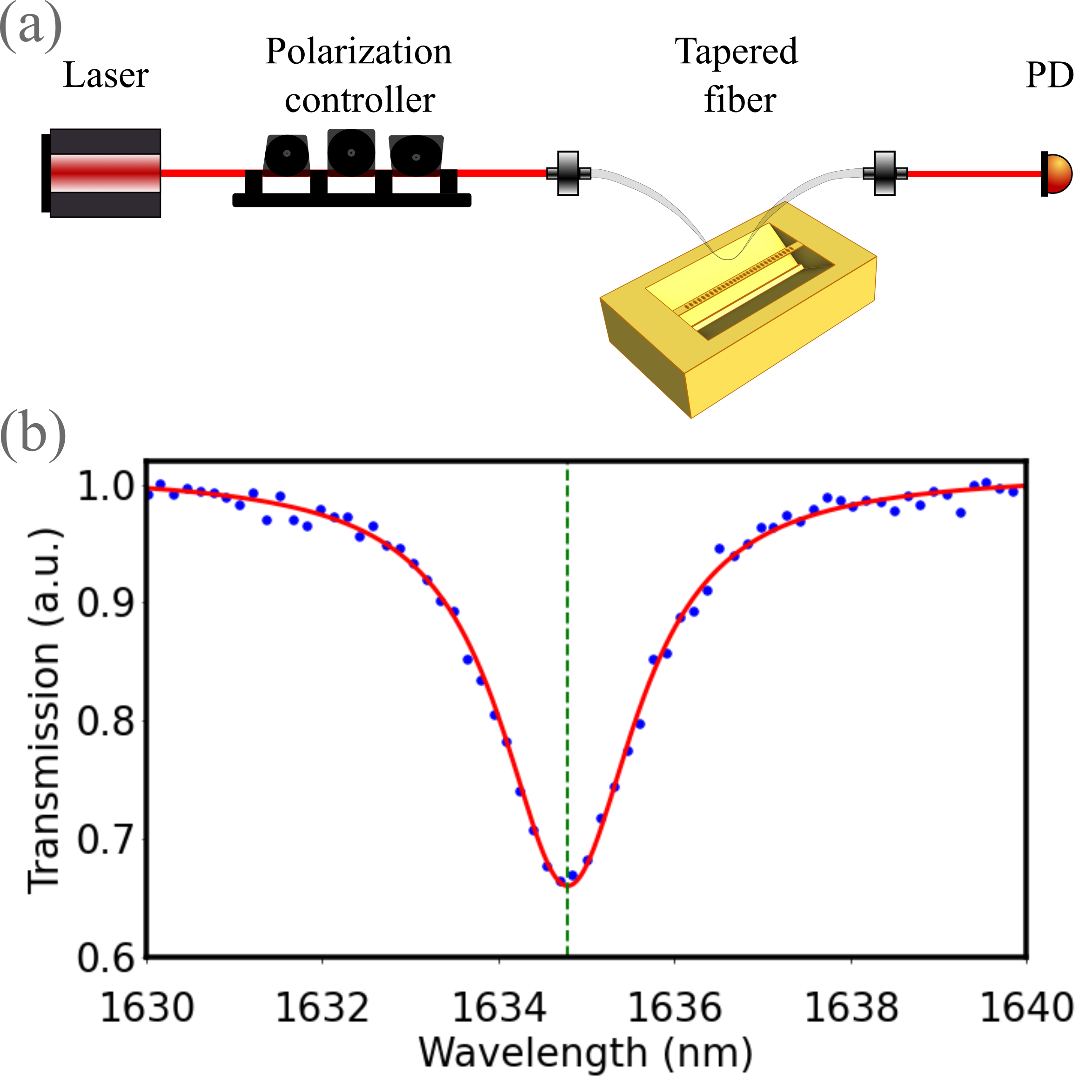}
\captionsetup{justification=justified}
\caption{\justifying \small (a) Schematic of the optical characterization setup. The setup uses a Santec TSL570 tunable laser source, with light from this source passing through a polarization controller to maximize coupling efficiency between the tapered dimpled fiber and the device. The transmitted signal is then detected by a Resolved Instruments DPD80 photodetector and analyzed. (b) The transmission spectrum reveals a resonance at $\lambda  = 1634.8 \text{ nm}$. By fitting a Lorentzian function to the data, we extract an internal quality factor of 2000 and an external coupling rate of $\kappa_{ex} = 9 \text{ GHz}$. The red solid line represents the Lorentzian fit, while the blue dots show the experimental data points.}
\label{Fig:03}
\end{figure}

After sputtering a 50 nm thick aluminum layer onto the YIG substrate, we milled two trenches to form a bridge structure, as shown in Fig.~\ref{Fig:02}(b). This initial milling creates the general outline of the device. Next, we carefully rotate the sample by 40 degrees to allow for an angled undercut of the bridge, forming a suspended triangular beam that will serve as the base of the waveguide, as illustrated in Fig.~\ref{Fig:02}(c). Following this, we switch to lower ion beam currents to finely trim and shape the beam down to its final dimensions: a width of \qty{1.25}{\micro \metre} and a length of \qty{20}{\micro \metre}. This reduction in current helps achieve greater precision, preventing unnecessary material removal and minimizing heat-related deformation. Once the beam has been trimmed, we return the sample to the default position (zero angle) and use a low-current ion beam to carefully drill the elliptical holes required for the design. These holes play a crucial role in defining the optical, mechanical and magnonic properties of the waveguide, so this step requires meticulous control to ensure their placement and size are accurate. The final step involves removing the sacrificial aluminum layer using a potassium hydroxide (KOH) bath. A 32\% KOH solution is prepared by diluting 45\% stock solution with deionized water. The bath is then heated to 80$^\circ$C, and the sample is immersed for 10 minutes to effectively etch away the thin aluminum layer. Following the etch, the sample is thoroughly rinsed with water and isopropanol (IPA) to remove any residual chemicals, leaving the YIG structure clean and free from contaminants.

Finally, the chip is dried using a critical point drying technique \cite{Horridge1969} with the Tousimis Autosamdri 815B system. After the KOH bath, the sample is kept in isopropanol to prevent premature drying. The device is then transferred into the dryer's chamber. Following the standard protocol, the chamber is purged for 10–20 minutes to remove residual IPA, then filled with liquid CO$_2$ and heated to achieve supercritical conditions. The chamber is subsequently depressurized in a controlled manner via a bleed and vent sequence. This process eliminates surface tension forces during drying, thereby preserving the integrity of the suspended nanostructure.

\section{Optical Characterization}

To characterize the optical properties, a tunable laser diode source directs light into an optical fiber polarization controller. The light is then guided into a tapered, dimpled optical fiber \cite{Hauer2014}, where the evanescent field of the dimpled region overlaps with the confined TE-like mode's field, enabling light coupling into and out of the device (Fig.~\ref{Fig:03}(a)). The light collected from the device by the dimpled fiber is detected by a Resolved Instruments 80 MHz photodetector and transmitted via USB for analysis.

The room temperature transmission spectrum of the OMC shows a resonance at $\lambda  = \qty{1634.8}{nm}$ with an internal quality factor of 2000 (Fig.~\ref{Fig:03}(b)), which is considerably lower than its simulated value of ${10}^6$. This discrepancy is attributed to surface roughness and fabrication imperfections, which can be addressed by optimizing the device geometry for FIB fabrication and will be addressed in the next generation of devices. This low internal quality factor (Q) places the device in the unresolved sideband regime \cite{Aspelmeyer2014}, defined by $\kappa \gg \Omega_m$, where $\Omega_m$ is the mechanical mode frequency and $\kappa$ is the optical mode decay rate. In this regime the optomechanical interaction is weak, which prevents an optical readout of the mechanical mode\cite{Kerdoncuff2015} and, consequently, blocks our planned optical readout of the magnon which relied on magnon–phonon coupling.  
\subsection{Origin of the optical quality factor discrepancy}
As noted above, the measured resonance at \(\lambda=\qty{1634.8}{nm}\) exhibits an intrinsic quality factor of \(\sim 2\times 10^3\), which is well below the simulated (radiation-limited) value of \(\sim 10^6\) (Fig.~\ref{Fig:03}(b)). Several mechanisms can contribute; here we summarize the ones most relevant to our device and identify the dominant source of loss.

First, focused-ion-beam (FIB) effects can modify the surface. Ga\(^+\) FIB is known to implant Ga and amorphize a thin surface layer, which increases absorption and scattering~\cite{Xiao2013,huang2015tridyn}. Our process employs Xe plasma FIB, which avoids Ga contamination and generally leads to less ion-species incorporation at comparable energies, although Xe can still produce a shallow damaged layer if milling parameters are not optimized~\cite{Hrn2015,Vitale2022}. To limit redeposition and implantation on the photonic surfaces we use a sacrificial Al layer during milling and remove it in wet etch, as detailed in Sec.~I.

Second, sidewall roughness and surface scattering add loss that ideal simulations neglect. At telecom wavelengths in high-index, small-mode-volume cavities, even a few-nanometer root-mean-square (rms) roughness can measurably limit Q; process improvements such as thorough cleans and thermal oxidation followed by oxide removal have raised experimental Q by orders of magnitude in silicon photonic-crystal nanocavities~\cite{Asano2017}. In our device, SEM images show tens-of-nanometer features along the sidewalls; while SEM does not quantify rms roughness, features at that scale are not negligible, so we regard roughness as a secondary, but nonzero, loss channel in this generation.

Finally, profile errors and lateral misalignment appear to be the dominant mechanism. Non-ideal etch profiles open radiation pathways that are absent in symmetric designs, and prior analyses of nanobeam photonic-crystal cavities identified the lateral offset of the air-hole row relative to the beam center as a leading cause of Q degradation~\cite{Yamaguchi2015}. In our device, SEM inspection indicates an offset of approximately \(\qty{100}{nm}\) together with a small deviation in the elliptical-hole aspect ratio. When the as-fabricated geometry is imported into COMSOL, the offset alone reduces the ideal Q by roughly two orders of magnitude, from \(\sim 10^6\) to \(\sim 10^4\). Including the aspect-ratio error further lowers the calculated Q and brings it close to the measured \(\sim 2\times 10^3\). We therefore attribute the low Q primarily to the lateral misplacement of the hole row, reinforced by the aspect-ratio deviation. To reduce sensitivity to such errors in the next iteration we will replace curved, sub-micron elliptical apertures with a straight-groove lattice across the beam, akin to geometries used in bulk-crystal nanophotonics~\cite{Zhong2016}, which desensitizes the cavity to lateral placement and has yielded high measured Q in the near-IR.

Looking ahead, we will focus on redesigning the photonic lattice for FIB transfer and improved alignment tolerance, guided by prior telecom-band demonstrations~\cite{Zhong2015,Zhong2016,Zhong2017}.

\section{Phononic and magnonic modes}
\subsection{Phononic mode}

\begin{figure}[b]
\includegraphics[width= 8.5 cm]{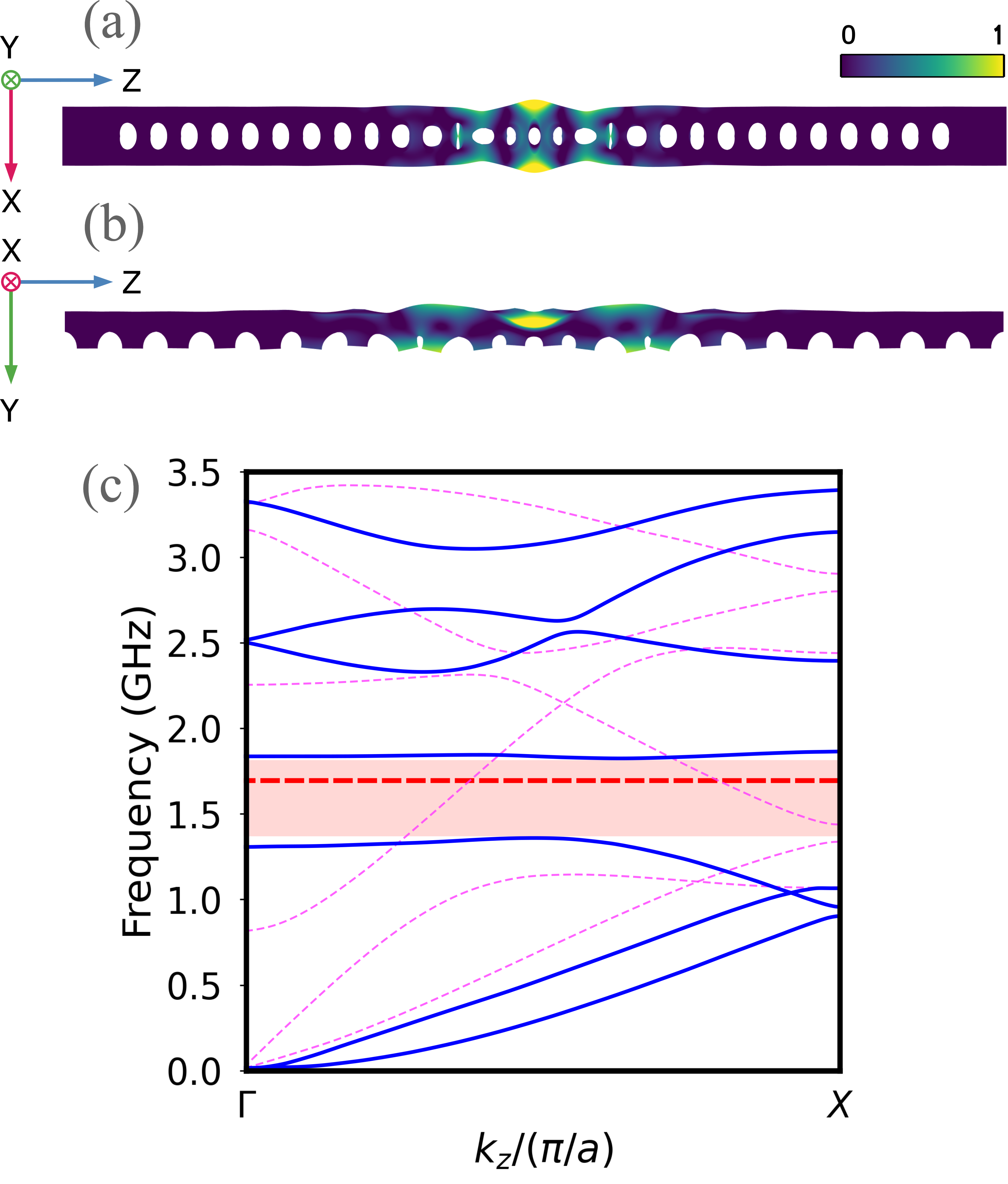}
\captionsetup{justification=justified}
\caption{\justifying \small (a),(b) Displacement field of the mechanical mode (not to scale) at $\Omega_{\,m} / 2\pi \approx \qty{1.52}{GHz}$. (c) Mechanical band diagram of the phononic waveguide. The $x$-symmetric and $x$-antisymmetric modes are represented by solid blue lines and dashed magenta lines, respectively. The bandgap formed by the $x$-symmetric modes is shaded in pink, with the supported confined mode indicated by the red horizontal dashed line.}
\label{Fig:04}
\end{figure}

We designed our photonic crystal with mechanical and magnon modes in mind, allowing the waveguide to function as a photonic, phononic, and magnonic waveguide, featuring overlapping guided photonic and phononic modes. The mechanical band diagram, shown in Fig.~\ref{Fig:04}(b), highlights a pink-shaded bandgap region. Similar to the photonic case, the phononic waveguide confines an $x$-symmetric phononic mode at $\Omega_{\,m} / 2\pi \approx \qty{1.52}{GHz}$ within this mechanical bandgap (Fig.~\ref{Fig:04}(b)). The mode profile and regions of maximum strain are illustrated in Fig.~\ref{Fig:04}(a), demonstrating the mode's potential for strong interactions with the photonic mode via radiation pressure.

To achieve this design, we conducted a series of finite element method (FEM) simulations in COMSOL Multiphysics to maximize the bandgaps and optimize both the optical and mechanical quality factors simultaneously  \cite{Chan2012}. Once the photonic and phononic modes were identified, we calculated the anticipated single photon-single phonon coupling rate, $g_0$ \cite{Chan2012}, to be \qty{50}{kHz}. While this value is low compared to counterparts fabricated in materials such as silicon and silicon nitride, it is attributed to YIG's relatively lower photoelastic tensor values \cite{Dixon1967, Lynch1973}. These values contribute to a reduced photoelastic coupling term, resulting in an order-of-magnitude decrease in coupling strength for our YIG OMC when compared to silicon-based devices \cite{Li2015, Chan2012, Benevides2017}.  To address this limitation, future iterations of the device will focus on a design where the moving boundary effect is the dominant coupling mechanism \cite{Li2015}, as seen in `zipper' and slot-mode optomechanical cavities \cite{Eichenfield2009_1, Chan2009, Leijssen2015}. This approach is expected to enhance the coupling efficiency by taking advantage of structural motion more effectively than the current design.

\subsection{Magnonic mode}

In addition to supporting photonic and phononic modes, our optomechanical crystal design facilitates the confinement and manipulation of magnonic modes within YIG. Similar to how photonic crystals control the propagation of light, magnonic crystals enable control over spin wave dynamics by introducing periodic variations in magnetic properties \cite{Graf2021}. By applying a strong external magnetic field, we align the magnetization in a specific direction, allowing for precise investigation of spin wave modes in our nanostructure.

Given the scale of our device, dipolar spin waves are the primary focus of our study \cite{Graf2021}. These spin waves are characterized based on their propagation direction relative to the magnetization vector $\vec{M}_0$ and the external magnetic field $\vec{H}_{\text{ext}}$. In our configuration, where the external magnetic field is applied in-plane along the $z$-axis, we excite backward volume spin waves, which are well-suited for our experimental objectives \cite{Venkat2013}.

To simulate the magnonic modes in our structure, we employ the finite-difference simulation tool MuMax3 \cite{Vansteenkiste2014}, which numerically solves the Landau-Lifshitz-Gilbert (LLG) equation for the local magnetization dynamics. We use material parameters specific to YIG: a saturation magnetization $M_s = \qty{140}{kA/m}$, an exchange stiffness constant $A_{\textrm{ex}} = \qty{2}{pJ/m}$, and a cubic anisotropy constant $K_{\textrm{c}} = \qty{-610}{J/m^3}$, with the anisotropy axis aligned along the $y$-axis \cite{Losby2015}. The external magnetic field is set to $H_{\textrm{ext}} = \qty{400}{mT}$.

\begin{figure}[t]
\includegraphics[width=8.5cm]{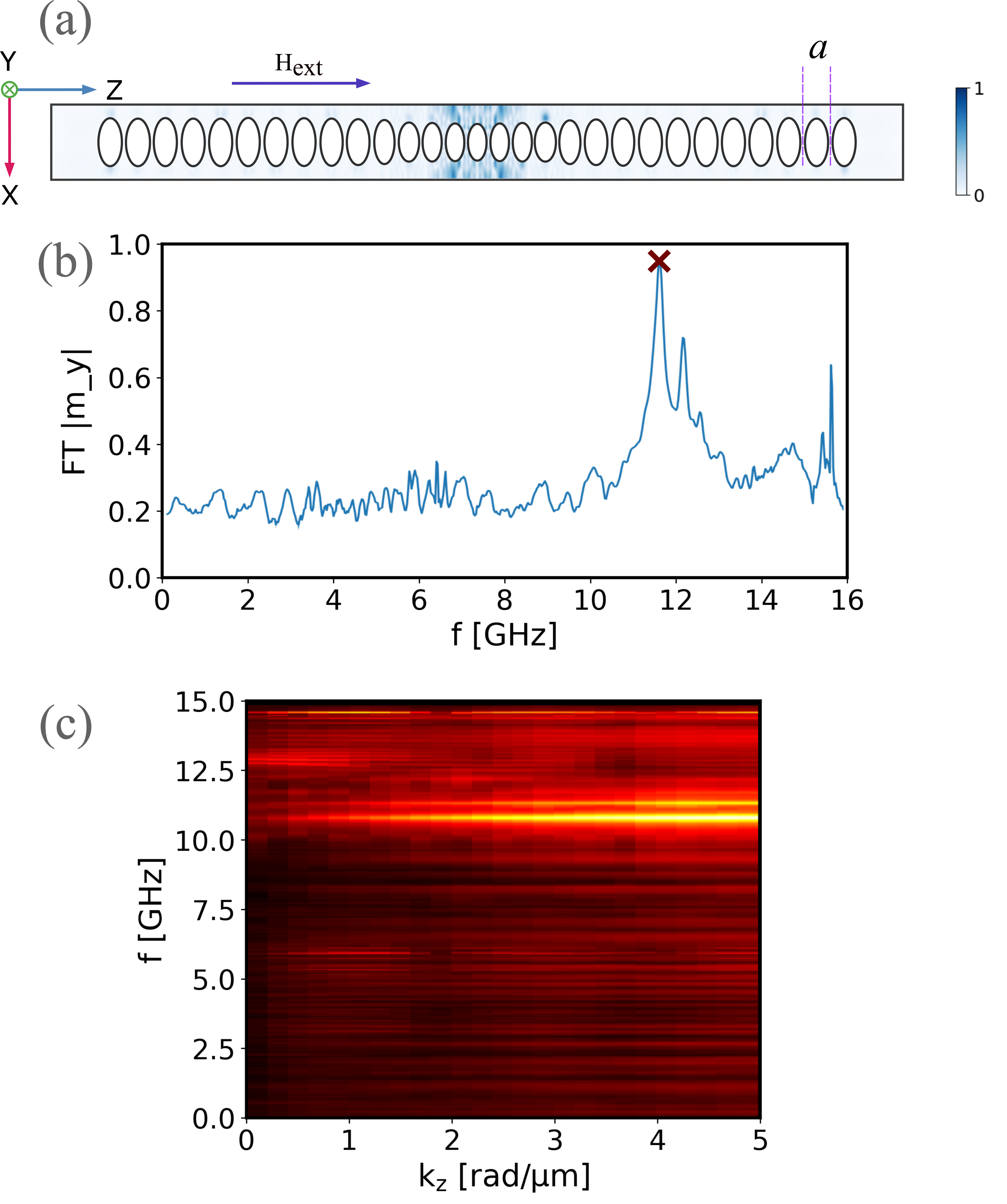}
\captionsetup{justification=justified}
\caption{\justifying \small Magnonic mode simulations. (a) Spatial profile of the $y$-component of the reduced magnetization ($|m_y|$) at $\omega_{\textrm{mag}} = \qty{11.59}{GHz}$, showing the confined magnonic mode within the structure. (b) Mode spectrum at the edge of the Brillouin zone ($k_z = \pi/a$, where $a = \qty{650}{nm}$ is the lattice constant), indicating the presence of the localized mode. (c) Band diagram of the backward volume waves within the Brillouin zone, illustrating the propagation characteristics of the spin waves in the magnonic crystal. It should be noted that, while the optical and mechanical modes were optimized using COMSOL, the magnonic modes have not undergone a similar optimization process. MuMax3 does not currently support the same parameterization strategies as COMSOL, limiting our ability to refine the geometry for maximum magnonic performance in this iteration.}
\label{Fig:05}
\end{figure}

Our simulations reveal that the magnonic crystal structure effectively confines specific magnonic modes within the device (Fig.~\ref{Fig:05}). A localized magnonic mode is predicted at a frequency of $\omega_{\text{mag}} = 2\pi \times \qty{11.59}{GHz}$. This confinement results from the periodic geometry of the structure and the in-plane applied magnetic field, which together tailor the spin wave dynamics to support backward volume waves.

Figure~\ref{Fig:05}(a) shows the spatial profile of the reduced magnetization's $y$-component ($|m_y|$, where $\vec{m}=\vec{M}/M_s$ with $M_s$ representing the saturation magnetization) for the confined magnonic mode. The highest intensity is concentrated in the central region of the device, demonstrating spatial localization. The band structure of the backward volume waves across the Brillouin zone is presented in Fig.~\ref{Fig:05}(c), illustrating the propagation of these waves within the crystal lattice. This co-localization is essential for exploring new phenomena in quantum hybrid systems and could lead to advancements in quantum information processing, such as enhanced magnon-based quantum memories and high-sensitivity magnetometry. 

As noted in Sec.~II, operation in the unresolved sideband regime prevented detection of the magnon mode. An alternative path is to characterize the magnon and the magnon-phonon coupling through cavity magnomechanics. Implementing an on-chip planar microwave resonator~\cite{Pitanti2015, Fink2016, Dieterle2016, Li2019, Verma2024, Bhoi2017} would provide a microwave port that could directly address the magnon and parametrically transduce its motion onto the mechanics, even when the optical cavity is not sideband resolved. This port would allow us to extract the magnon linewidth and resonance frequency, as well as the magnomechanical coupling rate \(g_{mb}\), while also supplying a redundant readout channel. In future work we will incorporate such a microwave resonator to enable cavity-magnomechanical measurements as we continue to raise the optical Q toward the sideband-resolved regime.

\section{Outlook}
Applications such as quantum wavelength transduction \cite{Engelhardt2022}, rely on the magnon-phonon interaction as an intermediate between optical and microwave photons. Nevertheless, accurately computing the magnon-phonon coupling rate presents significant challenges due to limitations in current simulation tools.

It is possible to estimate an upper limit on the magnomechanical coupling. The coupling between magnons and phonons can be derived from the quantization procedure used in Refs.~\cite{Gonzales2020,Bittencourt2023,Engelhardt2022}. Such procedure yields different types of interactions, in particular, potentially strong linear coupling, which is the one relevant for optical to microwave transduction \cite{Engelhardt2022}. The maximum achievable coupling rate is obtained by considering perfectly matching of specific combinations of mode functions \cite{Engelhardt2022}. Under this condition and using the parameters for YIG, we obtain for a phonon mode with frequency $\Omega_b/2 \pi = 1.52$ GHz a estimation $g_{mb}\sim 37$ MHz for the linear magnomechanical coupling. Even though such a coupling rate is large, the magnon and phonon modes in our current device are off-resonant. While the magnon frequency is tunable with the external field, our geometry requires an in-plane bias of \(H_{\mathrm{ext}}=400\,\mathrm{mT}\), which sets a lower bound near \(12~\mathrm{GHz}\) and permits only upward tuning. Future iterations of our YIG optomechanical crystal will focus on resonantly matching the mechanical and magnonic modes to enable stronger, linear interactions. In particular, pushing the mechanical mode frequencies into the 10--12\,GHz range places them closer to the magnonic spectrum, thus enhancing the single magnon-phonon coupling. In the off-resonant (parametric) regime, the single phonon-magnon coupling is typically orders of magnitude smaller but can be cavity-enhanced by applying a strong pump. While this approach offers unbounded enhancement to the single magnon-phonon coupling, it is ultimately constrained by nonlinearity and heating. By contrast, a resonant magnon-phonon interaction would reduce added noise and thermal load on the system by eliminating the need for the pump.

Nevertheless, operating in the parametric regime has advantages. Magnons in YIG are tunable via the external magnetic field, allowing the magnon frequency to shift by several tens of gigahertz. This flexibility is particularly valuable for transduction, as it can enable a universal transducer capable of interfacing with a variety of microwave quantum platforms, thus paving the way for a fully connected quantum network.

\section{Conclusion}

We have demonstrated the first nanofabricated air-suspended photonic crystal nanobeam cavity made from yttrium iron garnet, observing the intended optical resonance at a wavelength of $\lambda  = \qty{1634.8}{nm}$ with an internal quality factor of approximately 2000. This device is designed to simultaneously host optical photons, phonons, and magnons. This achievement represents an advancement in hybrid quantum systems, promising the co-localization and interaction of all three of the above quasiparticles within a shared nanoscale volume. It opens new avenues for quantum information processing and networking applications. Specifically, our work lays the foundation for developing efficient quantum transducers for coherent conversion between microwave and optical photons \cite{Engelhardt2022}, a crucial step toward interfacing quantum processors with optical communication networks \cite{Simon2017}. Additionally, our device has the potential to advance magnon-based quantum memories \cite{Zhang2015} and enable the integration of storage and transduction functionalities within a single device.

However, due to the relatively low optical quality factor in our current device, we were unable to resolve the mechanical mode in the present generation. This limitation highlights the need for further optimization of the device. Future work will focus on refining the fabrication process and device design to improve optical and mechanical quality factors, as well as to enhance the magnon-phonon and phonon-photon coupling rates. By addressing these challenges, we aim to realize functional optomagnomechanical devices that fulfill the potential envisioned in this work.

\section*{Acknowledgments}

We gratefully acknowledge financial support from the Natural Sciences and Engineering Research Council of Canada (NSERC), including RGPIN-2022-40378, CREATE-495446-17, ALLRP 558609–21, and ALLRP 597289–24. Additional support was provided by Alberta Innovates (AI), the National Research Council Canada (Quantum Sensing Program).

\section*{DATA AVAILABILITY}
The data that support the findings of this study are available from the corresponding author upon reasonable request.

\nocite{*}
\bibliographystyle{apsrev4-2}
\bibliography{apssamp}

\providecommand{\noopsort}[1]{}\providecommand{\singleletter}[1]{#1}%
\begin{thebibliography}{96}%
\makeatletter
\providecommand \@ifxundefined [1]{%
 \@ifx{#1\undefined}
}%
\providecommand \@ifnum [1]{%
 \ifnum #1\expandafter \@firstoftwo
 \else \expandafter \@secondoftwo
 \fi
}%
\providecommand \@ifx [1]{%
 \ifx #1\expandafter \@firstoftwo
 \else \expandafter \@secondoftwo
 \fi
}%
\providecommand \natexlab [1]{#1}%
\providecommand \enquote  [1]{``#1''}%
\providecommand \bibnamefont  [1]{#1}%
\providecommand \bibfnamefont [1]{#1}%
\providecommand \citenamefont [1]{#1}%
\providecommand \href@noop [0]{\@secondoftwo}%
\providecommand \href [0]{\begingroup \@sanitize@url \@href}%
\providecommand \@href[1]{\@@startlink{#1}\@@href}%
\providecommand \@@href[1]{\endgroup#1\@@endlink}%
\providecommand \@sanitize@url [0]{\catcode `\\12\catcode `\$12\catcode `\&12\catcode `\#12\catcode `\^12\catcode `\_12\catcode `\%12\relax}%
\providecommand \@@startlink[1]{}%
\providecommand \@@endlink[0]{}%
\providecommand \url  [0]{\begingroup\@sanitize@url \@url }%
\providecommand \@url [1]{\endgroup\@href {#1}{\urlprefix }}%
\providecommand \urlprefix  [0]{URL }%
\providecommand \Eprint [0]{\href }%
\providecommand \doibase [0]{https://doi.org/}%
\providecommand \selectlanguage [0]{\@gobble}%
\providecommand \bibinfo  [0]{\@secondoftwo}%
\providecommand \bibfield  [0]{\@secondoftwo}%
\providecommand \translation [1]{[#1]}%
\providecommand \BibitemOpen [0]{}%
\providecommand \bibitemStop [0]{}%
\providecommand \bibitemNoStop [0]{.\EOS\space}%
\providecommand \EOS [0]{\spacefactor3000\relax}%
\providecommand \BibitemShut  [1]{\csname bibitem#1\endcsname}%
\let\auto@bib@innerbib\@empty
\bibitem [{\citenamefont {Lachance-Quirion}\ \emph {et~al.}(2019)\citenamefont {Lachance-Quirion}, \citenamefont {Tabuchi}, \citenamefont {Gloppe}, \citenamefont {Usami},\ and\ \citenamefont {Nakamura}}]{Lachance-Quirion2019}%
  \BibitemOpen
  \bibfield  {author} {\bibinfo {author} {\bibfnamefont {D.}~\bibnamefont {Lachance-Quirion}}, \bibinfo {author} {\bibfnamefont {Y.}~\bibnamefont {Tabuchi}}, \bibinfo {author} {\bibfnamefont {A.}~\bibnamefont {Gloppe}}, \bibinfo {author} {\bibfnamefont {K.}~\bibnamefont {Usami}},\ and\ \bibinfo {author} {\bibfnamefont {Y.}~\bibnamefont {Nakamura}},\ }\href {https://doi.org/10.7567/1882-0786/ab248d} {\bibfield  {journal} {\bibinfo  {journal} {APEX}\ }\textbf {\bibinfo {volume} {12}},\ \bibinfo {pages} {070101} (\bibinfo {year} {2019})}\BibitemShut {NoStop}%
\bibitem [{\citenamefont {Aspelmeyer}\ \emph {et~al.}(2014)\citenamefont {Aspelmeyer}, \citenamefont {Kippenberg},\ and\ \citenamefont {Marquardt}}]{Aspelmeyer2014}%
  \BibitemOpen
  \bibfield  {author} {\bibinfo {author} {\bibfnamefont {M.}~\bibnamefont {Aspelmeyer}}, \bibinfo {author} {\bibfnamefont {T.~J.}\ \bibnamefont {Kippenberg}},\ and\ \bibinfo {author} {\bibfnamefont {F.}~\bibnamefont {Marquardt}},\ }\href {https://doi.org/10.1103/RevModPhys.86.1391} {\bibfield  {journal} {\bibinfo  {journal} {Rev. Mod. Phys.}\ }\textbf {\bibinfo {volume} {86}},\ \bibinfo {pages} {1391} (\bibinfo {year} {2014})}\BibitemShut {NoStop}%
\bibitem [{\citenamefont {Gröblacher}\ \emph {et~al.}(2009)\citenamefont {Gröblacher}, \citenamefont {Hammerer}, \citenamefont {Vanner},\ and\ \citenamefont {Aspelmeyer}}]{Gröblacher2009}%
  \BibitemOpen
  \bibfield  {author} {\bibinfo {author} {\bibfnamefont {S.}~\bibnamefont {Gröblacher}}, \bibinfo {author} {\bibfnamefont {K.}~\bibnamefont {Hammerer}}, \bibinfo {author} {\bibfnamefont {M.~R.}\ \bibnamefont {Vanner}},\ and\ \bibinfo {author} {\bibfnamefont {M.}~\bibnamefont {Aspelmeyer}},\ }\href {https://doi.org/10.1038/nature08171} {\bibfield  {journal} {\bibinfo  {journal} {Nature}\ }\textbf {\bibinfo {volume} {460}},\ \bibinfo {pages} {724} (\bibinfo {year} {2009})}\BibitemShut {NoStop}%
\bibitem [{\citenamefont {Barzanjeh}\ \emph {et~al.}(2022)\citenamefont {Barzanjeh}, \citenamefont {Xuereb}, \citenamefont {Gröblacher}, \citenamefont {Paternostro}, \citenamefont {Regal},\ and\ \citenamefont {Weig}}]{Barzanjeh2022}%
  \BibitemOpen
  \bibfield  {author} {\bibinfo {author} {\bibfnamefont {S.}~\bibnamefont {Barzanjeh}}, \bibinfo {author} {\bibfnamefont {A.}~\bibnamefont {Xuereb}}, \bibinfo {author} {\bibfnamefont {S.}~\bibnamefont {Gröblacher}}, \bibinfo {author} {\bibfnamefont {M.}~\bibnamefont {Paternostro}}, \bibinfo {author} {\bibfnamefont {C.~A.}\ \bibnamefont {Regal}},\ and\ \bibinfo {author} {\bibfnamefont {E.~M.}\ \bibnamefont {Weig}},\ }\href {https://doi.org/10.1038/s41567-021-01402-0} {\bibfield  {journal} {\bibinfo  {journal} {Nat. Phys.}\ }\textbf {\bibinfo {volume} {18}},\ \bibinfo {pages} {15} (\bibinfo {year} {2022})}\BibitemShut {NoStop}%
\bibitem [{\citenamefont {et~al}(2022)}]{Abbott2022}%
  \BibitemOpen
  \bibfield  {author} {\bibinfo {author} {\bibfnamefont {R.~A.}\ \bibnamefont {et~al}},\ }\href {https://doi.org/10.1103/PhysRevD.105.063030} {\bibfield  {journal} {\bibinfo  {journal} {Phys. Rev. D.}\ }\textbf {\bibinfo {volume} {105}},\ \bibinfo {pages} {063030} (\bibinfo {year} {2022})}\BibitemShut {NoStop}%
\bibitem [{\citenamefont {Manley}\ \emph {et~al.}(2020)\citenamefont {Manley}, \citenamefont {Wilson}, \citenamefont {Stump}, \citenamefont {Grin},\ and\ \citenamefont {Singh}}]{Manley2020}%
  \BibitemOpen
  \bibfield  {author} {\bibinfo {author} {\bibfnamefont {J.}~\bibnamefont {Manley}}, \bibinfo {author} {\bibfnamefont {D.~J.}\ \bibnamefont {Wilson}}, \bibinfo {author} {\bibfnamefont {R.}~\bibnamefont {Stump}}, \bibinfo {author} {\bibfnamefont {D.}~\bibnamefont {Grin}},\ and\ \bibinfo {author} {\bibfnamefont {S.}~\bibnamefont {Singh}},\ }\href {https://doi.org/10.1103/PhysRevLett.124.151301} {\bibfield  {journal} {\bibinfo  {journal} {PRL.}\ }\textbf {\bibinfo {volume} {124}},\ \bibinfo {pages} {151301} (\bibinfo {year} {2020})}\BibitemShut {NoStop}%
\bibitem [{\citenamefont {Baker}\ \emph {et~al.}(2024)\citenamefont {Baker}, \citenamefont {Bowen}, \citenamefont {Cox}, \citenamefont {Dolan}, \citenamefont {Goryachev},\ and\ \citenamefont {Harris}}]{Baker2024}%
  \BibitemOpen
  \bibfield  {author} {\bibinfo {author} {\bibfnamefont {C.~G.}\ \bibnamefont {Baker}}, \bibinfo {author} {\bibfnamefont {W.~P.}\ \bibnamefont {Bowen}}, \bibinfo {author} {\bibfnamefont {P.}~\bibnamefont {Cox}}, \bibinfo {author} {\bibfnamefont {M.~J.}\ \bibnamefont {Dolan}}, \bibinfo {author} {\bibfnamefont {M.}~\bibnamefont {Goryachev}},\ and\ \bibinfo {author} {\bibfnamefont {G.}~\bibnamefont {Harris}},\ }\href {https://doi.org/10.1103/PhysRevD.110.043005} {\bibfield  {journal} {\bibinfo  {journal} {Phys. Rev. D.}\ }\textbf {\bibinfo {volume} {110}},\ \bibinfo {pages} {043005} (\bibinfo {year} {2024})}\BibitemShut {NoStop}%
\bibitem [{\citenamefont {Vadakkumbatt}\ \emph {et~al.}(2021)\citenamefont {Vadakkumbatt}, \citenamefont {Hirschel}, \citenamefont {Manley}, \citenamefont {Clark}, \citenamefont {Singh},\ and\ \citenamefont {Davis}}]{Vadakkumbatt2021}%
  \BibitemOpen
  \bibfield  {author} {\bibinfo {author} {\bibfnamefont {V.}~\bibnamefont {Vadakkumbatt}}, \bibinfo {author} {\bibfnamefont {M.}~\bibnamefont {Hirschel}}, \bibinfo {author} {\bibfnamefont {J.}~\bibnamefont {Manley}}, \bibinfo {author} {\bibfnamefont {T.~J.}\ \bibnamefont {Clark}}, \bibinfo {author} {\bibfnamefont {S.}~\bibnamefont {Singh}},\ and\ \bibinfo {author} {\bibfnamefont {J.~P.}\ \bibnamefont {Davis}},\ }\href {https://doi.org/10.1103/PhysRevD.104.082001} {\bibfield  {journal} {\bibinfo  {journal} {Phys. Rev. D.}\ }\textbf {\bibinfo {volume} {104}},\ \bibinfo {pages} {082001} (\bibinfo {year} {2021})}\BibitemShut {NoStop}%
\bibitem [{\citenamefont {Hirschel}\ \emph {et~al.}(2024)\citenamefont {Hirschel}, \citenamefont {Vadakkumbatt}, \citenamefont {Baker}, \citenamefont {Schweizer}, \citenamefont {Sankey}, \citenamefont {Singh},\ and\ \citenamefont {Davis}}]{Hirschel2024}%
  \BibitemOpen
  \bibfield  {author} {\bibinfo {author} {\bibfnamefont {M.}~\bibnamefont {Hirschel}}, \bibinfo {author} {\bibfnamefont {V.}~\bibnamefont {Vadakkumbatt}}, \bibinfo {author} {\bibfnamefont {N.~P.}\ \bibnamefont {Baker}}, \bibinfo {author} {\bibfnamefont {F.~M.}\ \bibnamefont {Schweizer}}, \bibinfo {author} {\bibfnamefont {J.~C.}\ \bibnamefont {Sankey}}, \bibinfo {author} {\bibfnamefont {S.}~\bibnamefont {Singh}},\ and\ \bibinfo {author} {\bibfnamefont {J.~P.}\ \bibnamefont {Davis}},\ }\href {https://doi.org/10.1103/PhysRevD.109.095011} {\bibfield  {journal} {\bibinfo  {journal} {Phys. Rev. D.}\ }\textbf {\bibinfo {volume} {109}},\ \bibinfo {pages} {095011} (\bibinfo {year} {2024})}\BibitemShut {NoStop}%
\bibitem [{\citenamefont {Kim}\ \emph {et~al.}(2017)\citenamefont {Kim}, \citenamefont {Hauer}, \citenamefont {Clark}, \citenamefont {Sani}, \citenamefont {Freeman},\ and\ \citenamefont {Davis}}]{Kim2017}%
  \BibitemOpen
  \bibfield  {author} {\bibinfo {author} {\bibfnamefont {P.~H.}\ \bibnamefont {Kim}}, \bibinfo {author} {\bibfnamefont {B.~D.}\ \bibnamefont {Hauer}}, \bibinfo {author} {\bibfnamefont {T.~J.}\ \bibnamefont {Clark}}, \bibinfo {author} {\bibfnamefont {F.~F.}\ \bibnamefont {Sani}}, \bibinfo {author} {\bibfnamefont {M.~R.}\ \bibnamefont {Freeman}},\ and\ \bibinfo {author} {\bibfnamefont {J.~P.}\ \bibnamefont {Davis}},\ }\href {https://doi.org/10.1038/s41467-017-01380-z} {\bibfield  {journal} {\bibinfo  {journal} {Nat. Commun.}\ }\textbf {\bibinfo {volume} {8}},\ \bibinfo {pages} {1355} (\bibinfo {year} {2017})}\BibitemShut {NoStop}%
\bibitem [{\citenamefont {Rudd}\ \emph {et~al.}(2019)\citenamefont {Rudd}, \citenamefont {Kim}, \citenamefont {Potts}, \citenamefont {Doolin}, \citenamefont {Ramp}, \citenamefont {Hauer},\ and\ \citenamefont {Davis}}]{Rudd2019}%
  \BibitemOpen
  \bibfield  {author} {\bibinfo {author} {\bibfnamefont {M.~J.}\ \bibnamefont {Rudd}}, \bibinfo {author} {\bibfnamefont {P.~H.}\ \bibnamefont {Kim}}, \bibinfo {author} {\bibfnamefont {C.~A.}\ \bibnamefont {Potts}}, \bibinfo {author} {\bibfnamefont {C.}~\bibnamefont {Doolin}}, \bibinfo {author} {\bibfnamefont {H.}~\bibnamefont {Ramp}}, \bibinfo {author} {\bibfnamefont {B.~D.}\ \bibnamefont {Hauer}},\ and\ \bibinfo {author} {\bibfnamefont {J.~P.}\ \bibnamefont {Davis}},\ }\href {https://doi.org/10.1103/PhysRevApplied.12.034042} {\bibfield  {journal} {\bibinfo  {journal} {Phys. Rev. Appl.}\ }\textbf {\bibinfo {volume} {12}},\ \bibinfo {pages} {034042} (\bibinfo {year} {2019})}\BibitemShut {NoStop}%
\bibitem [{\citenamefont {Losby}\ \emph {et~al.}(2018)\citenamefont {Losby}, \citenamefont {Sauer},\ and\ \citenamefont {Freeman}}]{Losby2018}%
  \BibitemOpen
  \bibfield  {author} {\bibinfo {author} {\bibfnamefont {J.~E.}\ \bibnamefont {Losby}}, \bibinfo {author} {\bibfnamefont {V.~T.}\ \bibnamefont {Sauer}},\ and\ \bibinfo {author} {\bibfnamefont {M.~R.}\ \bibnamefont {Freeman}},\ }\href {https://doi.org/10.1088/1361-6463/aadccb} {\bibfield  {journal} {\bibinfo  {journal} {J. Phys. D: Appl. Phys.}\ }\textbf {\bibinfo {volume} {51}},\ \bibinfo {pages} {483001} (\bibinfo {year} {2018})}\BibitemShut {NoStop}%
\bibitem [{\citenamefont {Jiang}\ \emph {et~al.}(2020)\citenamefont {Jiang}, \citenamefont {Sarabalis}, \citenamefont {Dahmani}, \citenamefont {Patel}, \citenamefont {Mayor}, \citenamefont {McKenna}, \citenamefont {Laer},\ and\ \citenamefont {Safavi-Naeini}}]{Jiang2020}%
  \BibitemOpen
  \bibfield  {author} {\bibinfo {author} {\bibfnamefont {W.}~\bibnamefont {Jiang}}, \bibinfo {author} {\bibfnamefont {C.~J.}\ \bibnamefont {Sarabalis}}, \bibinfo {author} {\bibfnamefont {Y.~D.}\ \bibnamefont {Dahmani}}, \bibinfo {author} {\bibfnamefont {R.~N.}\ \bibnamefont {Patel}}, \bibinfo {author} {\bibfnamefont {F.~M.}\ \bibnamefont {Mayor}}, \bibinfo {author} {\bibfnamefont {T.~P.}\ \bibnamefont {McKenna}}, \bibinfo {author} {\bibfnamefont {R.~V.}\ \bibnamefont {Laer}},\ and\ \bibinfo {author} {\bibfnamefont {A.~H.}\ \bibnamefont {Safavi-Naeini}},\ }\href {https://doi.org/10.1038/s41467-020-14863-3} {\bibfield  {journal} {\bibinfo  {journal} {Nat. Commun.}\ }\textbf {\bibinfo {volume} {11}},\ \bibinfo {pages} {1166} (\bibinfo {year} {2020})}\BibitemShut {NoStop}%
\bibitem [{\citenamefont {Forsch}\ \emph {et~al.}(2020)\citenamefont {Forsch}, \citenamefont {Stockill}, \citenamefont {Wallucks}, \citenamefont {Marinković}, \citenamefont {Gärtner}, \citenamefont {Norte}, \citenamefont {van Otten}, \citenamefont {Fiore}, \citenamefont {Srinivasan},\ and\ \citenamefont {Gröblacher}}]{Forsch2020}%
  \BibitemOpen
  \bibfield  {author} {\bibinfo {author} {\bibfnamefont {M.}~\bibnamefont {Forsch}}, \bibinfo {author} {\bibfnamefont {R.}~\bibnamefont {Stockill}}, \bibinfo {author} {\bibfnamefont {A.}~\bibnamefont {Wallucks}}, \bibinfo {author} {\bibfnamefont {I.}~\bibnamefont {Marinković}}, \bibinfo {author} {\bibfnamefont {C.}~\bibnamefont {Gärtner}}, \bibinfo {author} {\bibfnamefont {R.~A.}\ \bibnamefont {Norte}}, \bibinfo {author} {\bibfnamefont {F.}~\bibnamefont {van Otten}}, \bibinfo {author} {\bibfnamefont {A.}~\bibnamefont {Fiore}}, \bibinfo {author} {\bibfnamefont {K.}~\bibnamefont {Srinivasan}},\ and\ \bibinfo {author} {\bibfnamefont {S.}~\bibnamefont {Gröblacher}},\ }\href {https://doi.org/10.1038/s41567-019-0673-7} {\bibfield  {journal} {\bibinfo  {journal} {Nat. Phys.}\ }\textbf {\bibinfo {volume} {16}},\ \bibinfo {pages} {69} (\bibinfo {year} {2020})}\BibitemShut {NoStop}%
\bibitem [{\citenamefont {Ramp}\ \emph {et~al.}(2020)\citenamefont {Ramp}, \citenamefont {Clark}, \citenamefont {Hauer}, \citenamefont {Doolin}, \citenamefont {Balram}, \citenamefont {Srinivasan},\ and\ \citenamefont {Davis}}]{Ramp2020}%
  \BibitemOpen
  \bibfield  {author} {\bibinfo {author} {\bibfnamefont {H.}~\bibnamefont {Ramp}}, \bibinfo {author} {\bibfnamefont {T.~J.}\ \bibnamefont {Clark}}, \bibinfo {author} {\bibfnamefont {B.~D.}\ \bibnamefont {Hauer}}, \bibinfo {author} {\bibfnamefont {C.}~\bibnamefont {Doolin}}, \bibinfo {author} {\bibfnamefont {K.~C.}\ \bibnamefont {Balram}}, \bibinfo {author} {\bibfnamefont {K.}~\bibnamefont {Srinivasan}},\ and\ \bibinfo {author} {\bibfnamefont {J.~P.}\ \bibnamefont {Davis}},\ }\href@noop {} {\bibfield  {journal} {\bibinfo  {journal} {Appl. Phys. Lett.}\ }\textbf {\bibinfo {volume} {116}} (\bibinfo {year} {2020})}\BibitemShut {NoStop}%
\bibitem [{\citenamefont {Arnold}\ \emph {et~al.}(2020)\citenamefont {Arnold}, \citenamefont {Wulf}, \citenamefont {Barzanjeh}, \citenamefont {Redchenko}, \citenamefont {Rueda}, \citenamefont {Hease}, \citenamefont {Hassani},\ and\ \citenamefont {Fink}}]{Arnold2020}%
  \BibitemOpen
  \bibfield  {author} {\bibinfo {author} {\bibfnamefont {G.}~\bibnamefont {Arnold}}, \bibinfo {author} {\bibfnamefont {M.}~\bibnamefont {Wulf}}, \bibinfo {author} {\bibfnamefont {S.}~\bibnamefont {Barzanjeh}}, \bibinfo {author} {\bibfnamefont {E.~S.}\ \bibnamefont {Redchenko}}, \bibinfo {author} {\bibfnamefont {A.}~\bibnamefont {Rueda}}, \bibinfo {author} {\bibfnamefont {W.~J.}\ \bibnamefont {Hease}}, \bibinfo {author} {\bibfnamefont {F.}~\bibnamefont {Hassani}},\ and\ \bibinfo {author} {\bibfnamefont {J.~M.}\ \bibnamefont {Fink}},\ }\href {https://doi.org/10.1038/s41467-020-18269-z} {\bibfield  {journal} {\bibinfo  {journal} {Nat. Commun.}\ }\textbf {\bibinfo {volume} {11}},\ \bibinfo {pages} {4460} (\bibinfo {year} {2020})}\BibitemShut {NoStop}%
\bibitem [{\citenamefont {Weaver}\ \emph {et~al.}(2024)\citenamefont {Weaver}, \citenamefont {Duivestein}, \citenamefont {Bernasconi}, \citenamefont {Scharmer}, \citenamefont {Lemang}, \citenamefont {Thiel}, \citenamefont {Hijazi}, \citenamefont {Hensen}, \citenamefont {Gröblacher},\ and\ \citenamefont {Stockill}}]{Weaver2024}%
  \BibitemOpen
  \bibfield  {author} {\bibinfo {author} {\bibfnamefont {M.~J.}\ \bibnamefont {Weaver}}, \bibinfo {author} {\bibfnamefont {P.}~\bibnamefont {Duivestein}}, \bibinfo {author} {\bibfnamefont {A.~C.}\ \bibnamefont {Bernasconi}}, \bibinfo {author} {\bibfnamefont {S.}~\bibnamefont {Scharmer}}, \bibinfo {author} {\bibfnamefont {M.}~\bibnamefont {Lemang}}, \bibinfo {author} {\bibfnamefont {T.~C.}\ \bibnamefont {Thiel}}, \bibinfo {author} {\bibfnamefont {F.}~\bibnamefont {Hijazi}}, \bibinfo {author} {\bibfnamefont {B.}~\bibnamefont {Hensen}}, \bibinfo {author} {\bibfnamefont {S.}~\bibnamefont {Gröblacher}},\ and\ \bibinfo {author} {\bibfnamefont {R.}~\bibnamefont {Stockill}},\ }\href {https://doi.org/10.1038/s41565-023-01515-y} {\bibfield  {journal} {\bibinfo  {journal} {Nat. Nanotechnol.}\ }\textbf {\bibinfo {volume} {19}},\ \bibinfo {pages} {166} (\bibinfo {year} {2024})}\BibitemShut {NoStop}%
\bibitem [{\citenamefont {Brubaker}\ \emph {et~al.}(2022)\citenamefont {Brubaker}, \citenamefont {Kindem}, \citenamefont {Urmey}, \citenamefont {Mittal}, \citenamefont {Delaney}, \citenamefont {Burns}, \citenamefont {Vissers}, \citenamefont {Lehnert},\ and\ \citenamefont {Regal}}]{Brubaker2022}%
  \BibitemOpen
  \bibfield  {author} {\bibinfo {author} {\bibfnamefont {B.~M.}\ \bibnamefont {Brubaker}}, \bibinfo {author} {\bibfnamefont {J.~M.}\ \bibnamefont {Kindem}}, \bibinfo {author} {\bibfnamefont {M.~D.}\ \bibnamefont {Urmey}}, \bibinfo {author} {\bibfnamefont {S.}~\bibnamefont {Mittal}}, \bibinfo {author} {\bibfnamefont {R.~D.}\ \bibnamefont {Delaney}}, \bibinfo {author} {\bibfnamefont {P.~S.}\ \bibnamefont {Burns}}, \bibinfo {author} {\bibfnamefont {M.~R.}\ \bibnamefont {Vissers}}, \bibinfo {author} {\bibfnamefont {K.~W.}\ \bibnamefont {Lehnert}},\ and\ \bibinfo {author} {\bibfnamefont {C.~A.}\ \bibnamefont {Regal}},\ }\href {https://doi.org/10.1103/PhysRevX.12.021062} {\bibfield  {journal} {\bibinfo  {journal} {Phys. Rev. X.}\ }\textbf {\bibinfo {volume} {12}},\ \bibinfo {pages} {021062} (\bibinfo {year} {2022})}\BibitemShut {NoStop}%
\bibitem [{\citenamefont {Safavi-Naeini}\ \emph {et~al.}(2013)\citenamefont {Safavi-Naeini}, \citenamefont {Gröblacher}, \citenamefont {Hill}, \citenamefont {Chan}, \citenamefont {Aspelmeyer},\ and\ \citenamefont {Painter}}]{Safavi-Naeini2013}%
  \BibitemOpen
  \bibfield  {author} {\bibinfo {author} {\bibfnamefont {A.~H.}\ \bibnamefont {Safavi-Naeini}}, \bibinfo {author} {\bibfnamefont {S.}~\bibnamefont {Gröblacher}}, \bibinfo {author} {\bibfnamefont {J.~T.}\ \bibnamefont {Hill}}, \bibinfo {author} {\bibfnamefont {J.}~\bibnamefont {Chan}}, \bibinfo {author} {\bibfnamefont {M.}~\bibnamefont {Aspelmeyer}},\ and\ \bibinfo {author} {\bibfnamefont {O.}~\bibnamefont {Painter}},\ }\href@noop {} {\bibfield  {journal} {\bibinfo  {journal} {Nature}\ }\textbf {\bibinfo {volume} {500}} (\bibinfo {year} {2013})}\BibitemShut {NoStop}%
\bibitem [{\citenamefont {Riedinger}\ \emph {et~al.}(2018)\citenamefont {Riedinger}, \citenamefont {Wallucks}, \citenamefont {Marinković}, \citenamefont {Löschnauer}, \citenamefont {Aspelmeyer}, \citenamefont {Hong},\ and\ \citenamefont {Gröblacher}}]{Riedinger2018}%
  \BibitemOpen
  \bibfield  {author} {\bibinfo {author} {\bibfnamefont {R.}~\bibnamefont {Riedinger}}, \bibinfo {author} {\bibfnamefont {A.}~\bibnamefont {Wallucks}}, \bibinfo {author} {\bibfnamefont {I.}~\bibnamefont {Marinković}}, \bibinfo {author} {\bibfnamefont {C.}~\bibnamefont {Löschnauer}}, \bibinfo {author} {\bibfnamefont {M.}~\bibnamefont {Aspelmeyer}}, \bibinfo {author} {\bibfnamefont {S.}~\bibnamefont {Hong}},\ and\ \bibinfo {author} {\bibfnamefont {S.}~\bibnamefont {Gröblacher}},\ }\href {https://doi.org/10.1038/s41586-018-0036-z} {\bibfield  {journal} {\bibinfo  {journal} {Nature}\ }\textbf {\bibinfo {volume} {556}},\ \bibinfo {pages} {473} (\bibinfo {year} {2018})}\BibitemShut {NoStop}%
\bibitem [{\citenamefont {Riedinger}\ \emph {et~al.}(2016)\citenamefont {Riedinger}, \citenamefont {Hong}, \citenamefont {Norte}, \citenamefont {Slater}, \citenamefont {Shang}, \citenamefont {Krause}, \citenamefont {Anant}, \citenamefont {Aspelmeyer},\ and\ \citenamefont {Gröblacher}}]{Riedinger2016}%
  \BibitemOpen
  \bibfield  {author} {\bibinfo {author} {\bibfnamefont {R.}~\bibnamefont {Riedinger}}, \bibinfo {author} {\bibfnamefont {S.}~\bibnamefont {Hong}}, \bibinfo {author} {\bibfnamefont {R.~A.}\ \bibnamefont {Norte}}, \bibinfo {author} {\bibfnamefont {J.~A.}\ \bibnamefont {Slater}}, \bibinfo {author} {\bibfnamefont {J.}~\bibnamefont {Shang}}, \bibinfo {author} {\bibfnamefont {A.~G.}\ \bibnamefont {Krause}}, \bibinfo {author} {\bibfnamefont {V.}~\bibnamefont {Anant}}, \bibinfo {author} {\bibfnamefont {M.}~\bibnamefont {Aspelmeyer}},\ and\ \bibinfo {author} {\bibfnamefont {S.}~\bibnamefont {Gröblacher}},\ }\href {https://doi.org/10.1038/nature16536} {\bibfield  {journal} {\bibinfo  {journal} {Nature}\ }\textbf {\bibinfo {volume} {530}},\ \bibinfo {pages} {313} (\bibinfo {year} {2016})}\BibitemShut {NoStop}%
\bibitem [{\citenamefont {Hong}\ \emph {et~al.}(2017)\citenamefont {Hong}, \citenamefont {Riedinger}, \citenamefont {Marinković}, \citenamefont {Wallucks}, \citenamefont {Hofer}, \citenamefont {Norte}, \citenamefont {Aspelmeyer},\ and\ \citenamefont {Gröblacher}}]{Hong2017}%
  \BibitemOpen
  \bibfield  {author} {\bibinfo {author} {\bibfnamefont {S.}~\bibnamefont {Hong}}, \bibinfo {author} {\bibfnamefont {R.}~\bibnamefont {Riedinger}}, \bibinfo {author} {\bibfnamefont {I.}~\bibnamefont {Marinković}}, \bibinfo {author} {\bibfnamefont {A.}~\bibnamefont {Wallucks}}, \bibinfo {author} {\bibfnamefont {S.~G.}\ \bibnamefont {Hofer}}, \bibinfo {author} {\bibfnamefont {R.~A.}\ \bibnamefont {Norte}}, \bibinfo {author} {\bibfnamefont {M.}~\bibnamefont {Aspelmeyer}},\ and\ \bibinfo {author} {\bibfnamefont {S.}~\bibnamefont {Gröblacher}},\ }\href {https://doi.org/10.1126/science.aan7939} {\bibfield  {journal} {\bibinfo  {journal} {Science}\ }\textbf {\bibinfo {volume} {358}},\ \bibinfo {pages} {203} (\bibinfo {year} {2017})}\BibitemShut {NoStop}%
\bibitem [{\citenamefont {Wallucks}\ \emph {et~al.}(2020)\citenamefont {Wallucks}, \citenamefont {Marinković}, \citenamefont {Hensen}, \citenamefont {Stockill},\ and\ \citenamefont {Gröblacher}}]{Wallucks2020}%
  \BibitemOpen
  \bibfield  {author} {\bibinfo {author} {\bibfnamefont {A.}~\bibnamefont {Wallucks}}, \bibinfo {author} {\bibfnamefont {I.}~\bibnamefont {Marinković}}, \bibinfo {author} {\bibfnamefont {B.}~\bibnamefont {Hensen}}, \bibinfo {author} {\bibfnamefont {R.}~\bibnamefont {Stockill}},\ and\ \bibinfo {author} {\bibfnamefont {S.}~\bibnamefont {Gröblacher}},\ }\href {https://doi.org/10.1038/s41567-020-0891-z} {\bibfield  {journal} {\bibinfo  {journal} {Nat. Phys.}\ }\textbf {\bibinfo {volume} {16}},\ \bibinfo {pages} {772} (\bibinfo {year} {2020})}\BibitemShut {NoStop}%
\bibitem [{\citenamefont {Smith}\ \emph {et~al.}(2023)\citenamefont {Smith}, \citenamefont {Babaei}, \citenamefont {Narayanan},\ and\ \citenamefont {LeBlanc}}]{Smith2023}%
  \BibitemOpen
  \bibfield  {author} {\bibinfo {author} {\bibfnamefont {B.~D.}\ \bibnamefont {Smith}}, \bibinfo {author} {\bibfnamefont {B.}~\bibnamefont {Babaei}}, \bibinfo {author} {\bibfnamefont {A.}~\bibnamefont {Narayanan}},\ and\ \bibinfo {author} {\bibfnamefont {L.~J.}\ \bibnamefont {LeBlanc}},\ }\href {https://doi.org/10.1038/s42005-023-01455-y} {\bibfield  {journal} {\bibinfo  {journal} {Commun. Phys.}\ }\textbf {\bibinfo {volume} {6}},\ \bibinfo {pages} {338} (\bibinfo {year} {2023})}\BibitemShut {NoStop}%
\bibitem [{\citenamefont {Tu}\ \emph {et~al.}(2022)\citenamefont {Tu}, \citenamefont {Liao}, \citenamefont {Zhang}, \citenamefont {Liu}, \citenamefont {Zheng}, \citenamefont {Yang}, \citenamefont {Zhang}, \citenamefont {Yan},\ and\ \citenamefont {Zhu}}]{Tu2022}%
  \BibitemOpen
  \bibfield  {author} {\bibinfo {author} {\bibfnamefont {H.~T.}\ \bibnamefont {Tu}}, \bibinfo {author} {\bibfnamefont {K.~Y.}\ \bibnamefont {Liao}}, \bibinfo {author} {\bibfnamefont {Z.~X.}\ \bibnamefont {Zhang}}, \bibinfo {author} {\bibfnamefont {X.~H.}\ \bibnamefont {Liu}}, \bibinfo {author} {\bibfnamefont {S.~Y.}\ \bibnamefont {Zheng}}, \bibinfo {author} {\bibfnamefont {S.~Z.}\ \bibnamefont {Yang}}, \bibinfo {author} {\bibfnamefont {X.~D.}\ \bibnamefont {Zhang}}, \bibinfo {author} {\bibfnamefont {H.}~\bibnamefont {Yan}},\ and\ \bibinfo {author} {\bibfnamefont {S.~L.}\ \bibnamefont {Zhu}},\ }\href {https://doi.org/10.1038/s41566-022-00959-3} {\bibfield  {journal} {\bibinfo  {journal} {Nat. Photon.}\ }\textbf {\bibinfo {volume} {16}},\ \bibinfo {pages} {291} (\bibinfo {year} {2022})}\BibitemShut {NoStop}%
\bibitem [{\citenamefont {Eichenfield}\ \emph {et~al.}(2009{\natexlab{a}})\citenamefont {Eichenfield}, \citenamefont {Chan}, \citenamefont {Camacho}, \citenamefont {Vahala},\ and\ \citenamefont {Painter}}]{Eichenfield2009}%
  \BibitemOpen
  \bibfield  {author} {\bibinfo {author} {\bibfnamefont {M.}~\bibnamefont {Eichenfield}}, \bibinfo {author} {\bibfnamefont {J.}~\bibnamefont {Chan}}, \bibinfo {author} {\bibfnamefont {R.~M.}\ \bibnamefont {Camacho}}, \bibinfo {author} {\bibfnamefont {K.~J.}\ \bibnamefont {Vahala}},\ and\ \bibinfo {author} {\bibfnamefont {O.}~\bibnamefont {Painter}},\ }\href {https://doi.org/10.1038/nature08524} {\bibfield  {journal} {\bibinfo  {journal} {Nature}\ }\textbf {\bibinfo {volume} {462}},\ \bibinfo {pages} {78} (\bibinfo {year} {2009}{\natexlab{a}})}\BibitemShut {NoStop}%
\bibitem [{\citenamefont {Safavi-Naeini}\ \emph {et~al.}(2014)\citenamefont {Safavi-Naeini}, \citenamefont {Hill}, \citenamefont {Meenehan}, \citenamefont {Chan}, \citenamefont {Gröblacher},\ and\ \citenamefont {Painter}}]{Safavi-Naeini2014}%
  \BibitemOpen
  \bibfield  {author} {\bibinfo {author} {\bibfnamefont {A.~H.}\ \bibnamefont {Safavi-Naeini}}, \bibinfo {author} {\bibfnamefont {J.~T.}\ \bibnamefont {Hill}}, \bibinfo {author} {\bibfnamefont {S.}~\bibnamefont {Meenehan}}, \bibinfo {author} {\bibfnamefont {J.}~\bibnamefont {Chan}}, \bibinfo {author} {\bibfnamefont {S.}~\bibnamefont {Gröblacher}},\ and\ \bibinfo {author} {\bibfnamefont {O.}~\bibnamefont {Painter}},\ }\href {https://doi.org/10.1103/PhysRevLett.112.153603} {\bibfield  {journal} {\bibinfo  {journal} {PRL.}\ }\textbf {\bibinfo {volume} {112}},\ \bibinfo {pages} {153603} (\bibinfo {year} {2014})}\BibitemShut {NoStop}%
\bibitem [{\citenamefont {Ren}\ \emph {et~al.}(2020)\citenamefont {Ren}, \citenamefont {Matheny}, \citenamefont {MacCabe}, \citenamefont {Luo}, \citenamefont {Pfeifer}, \citenamefont {Mirhosseini},\ and\ \citenamefont {Painter}}]{Ren2020}%
  \BibitemOpen
  \bibfield  {author} {\bibinfo {author} {\bibfnamefont {H.}~\bibnamefont {Ren}}, \bibinfo {author} {\bibfnamefont {M.~H.}\ \bibnamefont {Matheny}}, \bibinfo {author} {\bibfnamefont {G.~S.}\ \bibnamefont {MacCabe}}, \bibinfo {author} {\bibfnamefont {J.}~\bibnamefont {Luo}}, \bibinfo {author} {\bibfnamefont {H.}~\bibnamefont {Pfeifer}}, \bibinfo {author} {\bibfnamefont {M.}~\bibnamefont {Mirhosseini}},\ and\ \bibinfo {author} {\bibfnamefont {O.}~\bibnamefont {Painter}},\ }\href {https://doi.org/10.1038/s41467-020-17182-9} {\bibfield  {journal} {\bibinfo  {journal} {Nat. Commun.}\ }\textbf {\bibinfo {volume} {11}},\ \bibinfo {pages} {3373} (\bibinfo {year} {2020})}\BibitemShut {NoStop}%
\bibitem [{\citenamefont {Benevides}\ \emph {et~al.}(2017)\citenamefont {Benevides}, \citenamefont {Santos}, \citenamefont {Luiz}, \citenamefont {Wiederhecker},\ and\ \citenamefont {Alegre}}]{Benevides2017}%
  \BibitemOpen
  \bibfield  {author} {\bibinfo {author} {\bibfnamefont {R.}~\bibnamefont {Benevides}}, \bibinfo {author} {\bibfnamefont {F.~G.}\ \bibnamefont {Santos}}, \bibinfo {author} {\bibfnamefont {G.~O.}\ \bibnamefont {Luiz}}, \bibinfo {author} {\bibfnamefont {G.~S.}\ \bibnamefont {Wiederhecker}},\ and\ \bibinfo {author} {\bibfnamefont {T.~P.~M.}\ \bibnamefont {Alegre}},\ }\href {https://doi.org/10.1038/s41598-017-02515-4} {\bibfield  {journal} {\bibinfo  {journal} {Sci. Rep.}\ }\textbf {\bibinfo {volume} {7}},\ \bibinfo {pages} {2491} (\bibinfo {year} {2017})}\BibitemShut {NoStop}%
\bibitem [{\citenamefont {Gomis-Bresco}\ \emph {et~al.}(2014)\citenamefont {Gomis-Bresco}, \citenamefont {Navarro-Urrios}, \citenamefont {Oudich}, \citenamefont {El-Jallal}, \citenamefont {Griol}, \citenamefont {Puerto}, \citenamefont {Chavez}, \citenamefont {Pennec}, \citenamefont {Djafari-Rouhani}, \citenamefont {Alzina}, \citenamefont {Martínez},\ and\ \citenamefont {Torres}}]{Gomis-Bresco2014}%
  \BibitemOpen
  \bibfield  {author} {\bibinfo {author} {\bibfnamefont {J.}~\bibnamefont {Gomis-Bresco}}, \bibinfo {author} {\bibfnamefont {D.}~\bibnamefont {Navarro-Urrios}}, \bibinfo {author} {\bibfnamefont {M.}~\bibnamefont {Oudich}}, \bibinfo {author} {\bibfnamefont {S.}~\bibnamefont {El-Jallal}}, \bibinfo {author} {\bibfnamefont {A.}~\bibnamefont {Griol}}, \bibinfo {author} {\bibfnamefont {D.}~\bibnamefont {Puerto}}, \bibinfo {author} {\bibfnamefont {E.}~\bibnamefont {Chavez}}, \bibinfo {author} {\bibfnamefont {Y.}~\bibnamefont {Pennec}}, \bibinfo {author} {\bibfnamefont {B.}~\bibnamefont {Djafari-Rouhani}}, \bibinfo {author} {\bibfnamefont {F.}~\bibnamefont {Alzina}}, \bibinfo {author} {\bibfnamefont {A.}~\bibnamefont {Martínez}},\ and\ \bibinfo {author} {\bibfnamefont {C.~M.}\ \bibnamefont {Torres}},\ }\href {https://doi.org/10.1038/ncomms5452} {\bibfield  {journal} {\bibinfo  {journal} {Nat. Commun.}\ }\textbf {\bibinfo {volume} {5}},\ \bibinfo {pages} {4452} (\bibinfo {year} {2014})}\BibitemShut {NoStop}%
\bibitem [{\citenamefont {Safavi-Naeini}\ \emph {et~al.}(2011)\citenamefont {Safavi-Naeini}, \citenamefont {Alegre}, \citenamefont {Chan}, \citenamefont {Eichenfield}, \citenamefont {Winger}, \citenamefont {Lin}, \citenamefont {Hill}, \citenamefont {Chang},\ and\ \citenamefont {Painter}}]{Amir2011}%
  \BibitemOpen
  \bibfield  {author} {\bibinfo {author} {\bibfnamefont {A.~H.}\ \bibnamefont {Safavi-Naeini}}, \bibinfo {author} {\bibfnamefont {T.~P.}\ \bibnamefont {Alegre}}, \bibinfo {author} {\bibfnamefont {J.}~\bibnamefont {Chan}}, \bibinfo {author} {\bibfnamefont {M.}~\bibnamefont {Eichenfield}}, \bibinfo {author} {\bibfnamefont {M.}~\bibnamefont {Winger}}, \bibinfo {author} {\bibfnamefont {Q.}~\bibnamefont {Lin}}, \bibinfo {author} {\bibfnamefont {J.~T.}\ \bibnamefont {Hill}}, \bibinfo {author} {\bibfnamefont {D.~E.}\ \bibnamefont {Chang}},\ and\ \bibinfo {author} {\bibfnamefont {O.}~\bibnamefont {Painter}},\ }\href {https://doi.org/10.1038/nature09933} {\bibfield  {journal} {\bibinfo  {journal} {Nature}\ }\textbf {\bibinfo {volume} {472}},\ \bibinfo {pages} {69} (\bibinfo {year} {2011})}\BibitemShut {NoStop}%
\bibitem [{\citenamefont {Grutter}\ \emph {et~al.}(2015)\citenamefont {Grutter}, \citenamefont {Davanco},\ and\ \citenamefont {Srinivasan}}]{Grutter2015}%
  \BibitemOpen
  \bibfield  {author} {\bibinfo {author} {\bibfnamefont {K.~E.}\ \bibnamefont {Grutter}}, \bibinfo {author} {\bibfnamefont {M.}~\bibnamefont {Davanco}},\ and\ \bibinfo {author} {\bibfnamefont {K.}~\bibnamefont {Srinivasan}},\ }\href {https://doi.org/10.1109/JSTQE.2014.2376966} {\bibfield  {journal} {\bibinfo  {journal} {JSTQE.}\ }\textbf {\bibinfo {volume} {21}},\ \bibinfo {pages} {61} (\bibinfo {year} {2015})}\BibitemShut {NoStop}%
\bibitem [{\citenamefont {Burek}\ \emph {et~al.}(2016)\citenamefont {Burek}, \citenamefont {Cohen}, \citenamefont {Meenehan}, \citenamefont {El-Sawah}, \citenamefont {Chia}, \citenamefont {Ruelle}, \citenamefont {Meesala}, \citenamefont {Rochman}, \citenamefont {Atikian}, \citenamefont {Markham}, \citenamefont {Twitchen}, \citenamefont {Lukin}, \citenamefont {Painter},\ and\ \citenamefont {Lončar}}]{Burek2016}%
  \BibitemOpen
  \bibfield  {author} {\bibinfo {author} {\bibfnamefont {M.~J.}\ \bibnamefont {Burek}}, \bibinfo {author} {\bibfnamefont {J.~D.}\ \bibnamefont {Cohen}}, \bibinfo {author} {\bibfnamefont {S.~M.}\ \bibnamefont {Meenehan}}, \bibinfo {author} {\bibfnamefont {N.}~\bibnamefont {El-Sawah}}, \bibinfo {author} {\bibfnamefont {C.}~\bibnamefont {Chia}}, \bibinfo {author} {\bibfnamefont {T.}~\bibnamefont {Ruelle}}, \bibinfo {author} {\bibfnamefont {S.}~\bibnamefont {Meesala}}, \bibinfo {author} {\bibfnamefont {J.}~\bibnamefont {Rochman}}, \bibinfo {author} {\bibfnamefont {H.~A.}\ \bibnamefont {Atikian}}, \bibinfo {author} {\bibfnamefont {M.}~\bibnamefont {Markham}}, \bibinfo {author} {\bibfnamefont {D.~J.}\ \bibnamefont {Twitchen}}, \bibinfo {author} {\bibfnamefont {M.~D.}\ \bibnamefont {Lukin}}, \bibinfo {author} {\bibfnamefont {O.}~\bibnamefont {Painter}},\ and\ \bibinfo {author} {\bibfnamefont {M.}~\bibnamefont {Lončar}},\ }\href {https://doi.org/10.1364/optica.3.001404} {\bibfield  {journal} {\bibinfo  {journal}
  {Optica}\ }\textbf {\bibinfo {volume} {3}},\ \bibinfo {pages} {1404} (\bibinfo {year} {2016})}\BibitemShut {NoStop}%
\bibitem [{\citenamefont {Zuo}\ \emph {et~al.}(2024)\citenamefont {Zuo}, \citenamefont {Fan}, \citenamefont {Qian}, \citenamefont {Ding}, \citenamefont {Tan}, \citenamefont {Xiong},\ and\ \citenamefont {Li}}]{Zuo2024}%
  \BibitemOpen
  \bibfield  {author} {\bibinfo {author} {\bibfnamefont {X.}~\bibnamefont {Zuo}}, \bibinfo {author} {\bibfnamefont {Z.~Y.}\ \bibnamefont {Fan}}, \bibinfo {author} {\bibfnamefont {H.}~\bibnamefont {Qian}}, \bibinfo {author} {\bibfnamefont {M.~S.}\ \bibnamefont {Ding}}, \bibinfo {author} {\bibfnamefont {H.}~\bibnamefont {Tan}}, \bibinfo {author} {\bibfnamefont {H.}~\bibnamefont {Xiong}},\ and\ \bibinfo {author} {\bibfnamefont {J.}~\bibnamefont {Li}},\ }\href {https://doi.org/10.1088/1367-2630/ad327c} {\bibfield  {journal} {\bibinfo  {journal} {New J. Phys.}\ }\textbf {\bibinfo {volume} {26}},\ \bibinfo {pages} {031201} (\bibinfo {year} {2024})}\BibitemShut {NoStop}%
\bibitem [{\citenamefont {Serga}\ \emph {et~al.}(2010)\citenamefont {Serga}, \citenamefont {Chumak},\ and\ \citenamefont {Hillebrands}}]{Serga2010}%
  \BibitemOpen
  \bibfield  {author} {\bibinfo {author} {\bibfnamefont {A.~A.}\ \bibnamefont {Serga}}, \bibinfo {author} {\bibfnamefont {A.~V.}\ \bibnamefont {Chumak}},\ and\ \bibinfo {author} {\bibfnamefont {B.}~\bibnamefont {Hillebrands}},\ }\href {https://doi.org/10.1088/0022-3727/43/26/264002} {\bibfield  {journal} {\bibinfo  {journal} {J. Phys. D: Appl. Phys.}\ }\textbf {\bibinfo {volume} {43}},\ \bibinfo {pages} {264002} (\bibinfo {year} {2010})}\BibitemShut {NoStop}%
\bibitem [{\citenamefont {Klingler}\ \emph {et~al.}(2017)\citenamefont {Klingler}, \citenamefont {Maier-Flaig}, \citenamefont {Dubs}, \citenamefont {Surzhenko}, \citenamefont {Gross}, \citenamefont {Huebl}, \citenamefont {Goennenwein},\ and\ \citenamefont {Weiler}}]{Klingler2017}%
  \BibitemOpen
  \bibfield  {author} {\bibinfo {author} {\bibfnamefont {S.}~\bibnamefont {Klingler}}, \bibinfo {author} {\bibfnamefont {H.}~\bibnamefont {Maier-Flaig}}, \bibinfo {author} {\bibfnamefont {C.}~\bibnamefont {Dubs}}, \bibinfo {author} {\bibfnamefont {O.}~\bibnamefont {Surzhenko}}, \bibinfo {author} {\bibfnamefont {R.}~\bibnamefont {Gross}}, \bibinfo {author} {\bibfnamefont {H.}~\bibnamefont {Huebl}}, \bibinfo {author} {\bibfnamefont {S.~T.}\ \bibnamefont {Goennenwein}},\ and\ \bibinfo {author} {\bibfnamefont {M.}~\bibnamefont {Weiler}},\ }\href@noop {} {\bibfield  {journal} {\bibinfo  {journal} {Appl. Phys. Lett.}\ }\textbf {\bibinfo {volume} {110}} (\bibinfo {year} {2017})}\BibitemShut {NoStop}%
\bibitem [{\citenamefont {Wemple}\ \emph {et~al.}(1974)\citenamefont {Wemple}, \citenamefont {Blank}, \citenamefont {Seman},\ and\ \citenamefont {Biolsi}}]{Wemple1974}%
  \BibitemOpen
  \bibfield  {author} {\bibinfo {author} {\bibfnamefont {S.~H.}\ \bibnamefont {Wemple}}, \bibinfo {author} {\bibfnamefont {S.~L.}\ \bibnamefont {Blank}}, \bibinfo {author} {\bibfnamefont {J.~A.}\ \bibnamefont {Seman}},\ and\ \bibinfo {author} {\bibfnamefont {W.~A.}\ \bibnamefont {Biolsi}},\ }\href {https://doi.org/10.1103/PhysRevB.9.2134} {\bibfield  {journal} {\bibinfo  {journal} {Phys. Rev. B.}\ }\textbf {\bibinfo {volume} {9}},\ \bibinfo {pages} {2134} (\bibinfo {year} {1974})}\BibitemShut {NoStop}%
\bibitem [{\citenamefont {Potts}\ \emph {et~al.}(2020)\citenamefont {Potts}, \citenamefont {Bittencourt}, \citenamefont {Kusminskiy}, \citenamefont {Kusminskiy},\ and\ \citenamefont {Davis}}]{Potts2020}%
  \BibitemOpen
  \bibfield  {author} {\bibinfo {author} {\bibfnamefont {C.~A.}\ \bibnamefont {Potts}}, \bibinfo {author} {\bibfnamefont {V.~A.}\ \bibnamefont {Bittencourt}}, \bibinfo {author} {\bibfnamefont {S.~V.}\ \bibnamefont {Kusminskiy}}, \bibinfo {author} {\bibfnamefont {S.~V.}\ \bibnamefont {Kusminskiy}},\ and\ \bibinfo {author} {\bibfnamefont {J.~P.}\ \bibnamefont {Davis}},\ }\href {https://doi.org/10.1103/PhysRevApplied.13.064001} {\bibfield  {journal} {\bibinfo  {journal} {Phys. Rev. Appl.}\ }\textbf {\bibinfo {volume} {13}},\ \bibinfo {pages} {064001} (\bibinfo {year} {2020})}\BibitemShut {NoStop}%
\bibitem [{\citenamefont {Potts}\ \emph {et~al.}(2023)\citenamefont {Potts}, \citenamefont {Huang}, \citenamefont {Bittencourt}, \citenamefont {Kusminskiy},\ and\ \citenamefont {Davis}}]{Potts2023}%
  \BibitemOpen
  \bibfield  {author} {\bibinfo {author} {\bibfnamefont {C.~A.}\ \bibnamefont {Potts}}, \bibinfo {author} {\bibfnamefont {Y.}~\bibnamefont {Huang}}, \bibinfo {author} {\bibfnamefont {V.~A.}\ \bibnamefont {Bittencourt}}, \bibinfo {author} {\bibfnamefont {S.~V.}\ \bibnamefont {Kusminskiy}},\ and\ \bibinfo {author} {\bibfnamefont {J.~P.}\ \bibnamefont {Davis}},\ }\href {https://doi.org/10.1103/PhysRevB.107.L140405} {\bibfield  {journal} {\bibinfo  {journal} {Phys. Rev. B.}\ }\textbf {\bibinfo {volume} {107}},\ \bibinfo {pages} {L140405} (\bibinfo {year} {2023})}\BibitemShut {NoStop}%
\bibitem [{\citenamefont {Bittencourt}\ \emph {et~al.}(2023)\citenamefont {Bittencourt}, \citenamefont {Potts}, \citenamefont {Huang}, \citenamefont {Davis},\ and\ \citenamefont {Kusminskiy}}]{Bittencourt2023}%
  \BibitemOpen
  \bibfield  {author} {\bibinfo {author} {\bibfnamefont {V.~A.}\ \bibnamefont {Bittencourt}}, \bibinfo {author} {\bibfnamefont {C.~A.}\ \bibnamefont {Potts}}, \bibinfo {author} {\bibfnamefont {Y.}~\bibnamefont {Huang}}, \bibinfo {author} {\bibfnamefont {J.~P.}\ \bibnamefont {Davis}},\ and\ \bibinfo {author} {\bibfnamefont {S.~V.}\ \bibnamefont {Kusminskiy}},\ }\href {https://doi.org/10.1103/PhysRevB.107.144411} {\bibfield  {journal} {\bibinfo  {journal} {Phys. Rev. B.}\ }\textbf {\bibinfo {volume} {107}},\ \bibinfo {pages} {144411} (\bibinfo {year} {2023})}\BibitemShut {NoStop}%
\bibitem [{\citenamefont {Liu}\ \emph {et~al.}(2023)\citenamefont {Liu}, \citenamefont {Peng},\ and\ \citenamefont {Xiong}}]{Liu2023}%
  \BibitemOpen
  \bibfield  {author} {\bibinfo {author} {\bibfnamefont {Z.~X.}\ \bibnamefont {Liu}}, \bibinfo {author} {\bibfnamefont {J.}~\bibnamefont {Peng}},\ and\ \bibinfo {author} {\bibfnamefont {H.}~\bibnamefont {Xiong}},\ }\href {https://doi.org/10.1103/PhysRevA.107.053708} {\bibfield  {journal} {\bibinfo  {journal} {Phys. Rev. A.}\ }\textbf {\bibinfo {volume} {107}},\ \bibinfo {pages} {053708} (\bibinfo {year} {2023})}\BibitemShut {NoStop}%
\bibitem [{\citenamefont {Li}\ \emph {et~al.}(2021)\citenamefont {Li}, \citenamefont {Wang}, \citenamefont {Wu}, \citenamefont {Zhu},\ and\ \citenamefont {You}}]{Li2021}%
  \BibitemOpen
  \bibfield  {author} {\bibinfo {author} {\bibfnamefont {J.}~\bibnamefont {Li}}, \bibinfo {author} {\bibfnamefont {Y.~P.}\ \bibnamefont {Wang}}, \bibinfo {author} {\bibfnamefont {W.~J.}\ \bibnamefont {Wu}}, \bibinfo {author} {\bibfnamefont {S.~Y.}\ \bibnamefont {Zhu}},\ and\ \bibinfo {author} {\bibfnamefont {J.~Q.}\ \bibnamefont {You}},\ }\href {https://doi.org/10.1103/PRXQuantum.2.040344} {\bibfield  {journal} {\bibinfo  {journal} {PRX Quantum}\ }\textbf {\bibinfo {volume} {2}},\ \bibinfo {pages} {040344} (\bibinfo {year} {2021})}\BibitemShut {NoStop}%
\bibitem [{\citenamefont {Zhang}\ \emph {et~al.}(2016{\natexlab{a}})\citenamefont {Zhang}, \citenamefont {Zhu}, \citenamefont {Zou},\ and\ \citenamefont {Tang}}]{Zhang2016}%
  \BibitemOpen
  \bibfield  {author} {\bibinfo {author} {\bibfnamefont {X.}~\bibnamefont {Zhang}}, \bibinfo {author} {\bibfnamefont {N.}~\bibnamefont {Zhu}}, \bibinfo {author} {\bibfnamefont {C.~L.}\ \bibnamefont {Zou}},\ and\ \bibinfo {author} {\bibfnamefont {H.~X.}\ \bibnamefont {Tang}},\ }\href {https://doi.org/10.1103/PhysRevLett.117.123605} {\bibfield  {journal} {\bibinfo  {journal} {PRL.}\ }\textbf {\bibinfo {volume} {117}},\ \bibinfo {pages} {123605} (\bibinfo {year} {2016}{\natexlab{a}})}\BibitemShut {NoStop}%
\bibitem [{\citenamefont {Zhang}\ \emph {et~al.}(2014)\citenamefont {Zhang}, \citenamefont {Zou}, \citenamefont {Jiang},\ and\ \citenamefont {Tang}}]{Zhang2014}%
  \BibitemOpen
  \bibfield  {author} {\bibinfo {author} {\bibfnamefont {X.}~\bibnamefont {Zhang}}, \bibinfo {author} {\bibfnamefont {C.~L.}\ \bibnamefont {Zou}}, \bibinfo {author} {\bibfnamefont {L.}~\bibnamefont {Jiang}},\ and\ \bibinfo {author} {\bibfnamefont {H.~X.}\ \bibnamefont {Tang}},\ }\href {https://doi.org/10.1103/PhysRevLett.113.156401} {\bibfield  {journal} {\bibinfo  {journal} {PRL.}\ }\textbf {\bibinfo {volume} {113}},\ \bibinfo {pages} {156401} (\bibinfo {year} {2014})}\BibitemShut {NoStop}%
\bibitem [{\citenamefont {Zhang}\ \emph {et~al.}(2015)\citenamefont {Zhang}, \citenamefont {Zou}, \citenamefont {Zhu}, \citenamefont {Marquardt}, \citenamefont {Jiang},\ and\ \citenamefont {Tang}}]{Zhang2015}%
  \BibitemOpen
  \bibfield  {author} {\bibinfo {author} {\bibfnamefont {X.}~\bibnamefont {Zhang}}, \bibinfo {author} {\bibfnamefont {C.~L.}\ \bibnamefont {Zou}}, \bibinfo {author} {\bibfnamefont {N.}~\bibnamefont {Zhu}}, \bibinfo {author} {\bibfnamefont {F.}~\bibnamefont {Marquardt}}, \bibinfo {author} {\bibfnamefont {L.}~\bibnamefont {Jiang}},\ and\ \bibinfo {author} {\bibfnamefont {H.~X.}\ \bibnamefont {Tang}},\ }\href {https://doi.org/10.1038/ncomms9914} {\bibfield  {journal} {\bibinfo  {journal} {Nat. Commun.}\ }\textbf {\bibinfo {volume} {6}},\ \bibinfo {pages} {8914} (\bibinfo {year} {2015})}\BibitemShut {NoStop}%
\bibitem [{\citenamefont {Engelhardt}\ \emph {et~al.}(2022)\citenamefont {Engelhardt}, \citenamefont {Bittencourt}, \citenamefont {Huebl}, \citenamefont {Klein},\ and\ \citenamefont {Kusminskiy}}]{Engelhardt2022}%
  \BibitemOpen
  \bibfield  {author} {\bibinfo {author} {\bibfnamefont {F.}~\bibnamefont {Engelhardt}}, \bibinfo {author} {\bibfnamefont {V.~A.}\ \bibnamefont {Bittencourt}}, \bibinfo {author} {\bibfnamefont {H.}~\bibnamefont {Huebl}}, \bibinfo {author} {\bibfnamefont {O.}~\bibnamefont {Klein}},\ and\ \bibinfo {author} {\bibfnamefont {S.~V.}\ \bibnamefont {Kusminskiy}},\ }\href {https://doi.org/10.1103/PhysRevApplied.18.044059} {\bibfield  {journal} {\bibinfo  {journal} {Phys. Rev. Appl.}\ }\textbf {\bibinfo {volume} {18}},\ \bibinfo {pages} {044059} (\bibinfo {year} {2022})}\BibitemShut {NoStop}%
\bibitem [{\citenamefont {Simon}(2017)}]{Simon2017}%
  \BibitemOpen
  \bibfield  {author} {\bibinfo {author} {\bibfnamefont {C.}~\bibnamefont {Simon}},\ }\href {https://doi.org/10.1038/s41566-017-0032-0} {\bibfield  {journal} {\bibinfo  {journal} {Nat. Photon.}\ }\textbf {\bibinfo {volume} {11}},\ \bibinfo {pages} {678} (\bibinfo {year} {2017})}\BibitemShut {NoStop}%
\bibitem [{\citenamefont {Zhu}\ \emph {et~al.}(2022)\citenamefont {Zhu}, \citenamefont {Zhang}, \citenamefont {Han}, \citenamefont {Zou},\ and\ \citenamefont {Tang}}]{Zhu2022}%
  \BibitemOpen
  \bibfield  {author} {\bibinfo {author} {\bibfnamefont {N.}~\bibnamefont {Zhu}}, \bibinfo {author} {\bibfnamefont {X.}~\bibnamefont {Zhang}}, \bibinfo {author} {\bibfnamefont {X.}~\bibnamefont {Han}}, \bibinfo {author} {\bibfnamefont {C.~L.}\ \bibnamefont {Zou}},\ and\ \bibinfo {author} {\bibfnamefont {H.~X.}\ \bibnamefont {Tang}},\ }\href {https://doi.org/10.1103/PhysRevApplied.18.024046} {\bibfield  {journal} {\bibinfo  {journal} {Phys. Rev. Appl.}\ }\textbf {\bibinfo {volume} {18}},\ \bibinfo {pages} {024046} (\bibinfo {year} {2022})}\BibitemShut {NoStop}%
\bibitem [{\citenamefont {Trempler}\ \emph {et~al.}(2020)\citenamefont {Trempler}, \citenamefont {Dreyer}, \citenamefont {Geyer}, \citenamefont {Hauser}, \citenamefont {Woltersdorf},\ and\ \citenamefont {Schmidt}}]{Trempler2020}%
  \BibitemOpen
  \bibfield  {author} {\bibinfo {author} {\bibfnamefont {P.}~\bibnamefont {Trempler}}, \bibinfo {author} {\bibfnamefont {R.}~\bibnamefont {Dreyer}}, \bibinfo {author} {\bibfnamefont {P.}~\bibnamefont {Geyer}}, \bibinfo {author} {\bibfnamefont {C.}~\bibnamefont {Hauser}}, \bibinfo {author} {\bibfnamefont {G.}~\bibnamefont {Woltersdorf}},\ and\ \bibinfo {author} {\bibfnamefont {G.}~\bibnamefont {Schmidt}},\ }\href@noop {} {\bibfield  {journal} {\bibinfo  {journal} {Appl. Phys. Lett.}\ }\textbf {\bibinfo {volume} {117}},\ \bibinfo {pages} {232401} (\bibinfo {year} {2020})}\BibitemShut {NoStop}%
\bibitem [{\citenamefont {Schmidt}\ \emph {et~al.}(2020)\citenamefont {Schmidt}, \citenamefont {Hauser}, \citenamefont {Trempler}, \citenamefont {Paleschke},\ and\ \citenamefont {Papaioannou}}]{Schmidt2020}%
  \BibitemOpen
  \bibfield  {author} {\bibinfo {author} {\bibfnamefont {G.}~\bibnamefont {Schmidt}}, \bibinfo {author} {\bibfnamefont {C.}~\bibnamefont {Hauser}}, \bibinfo {author} {\bibfnamefont {P.}~\bibnamefont {Trempler}}, \bibinfo {author} {\bibfnamefont {M.}~\bibnamefont {Paleschke}},\ and\ \bibinfo {author} {\bibfnamefont {E.~T.}\ \bibnamefont {Papaioannou}},\ }\href {https://doi.org/10.1002/pssb.201900644} {\bibfield  {journal} {\bibinfo  {journal} {Phys. Status Solidi B Basic Res.}\ }\textbf {\bibinfo {volume} {257}},\ \bibinfo {pages} {1900644} (\bibinfo {year} {2020})}\BibitemShut {NoStop}%
\bibitem [{\citenamefont {Arisawa}\ \emph {et~al.}(2019)\citenamefont {Arisawa}, \citenamefont {Daimon}, \citenamefont {Oikawa}, \citenamefont {Seo}, \citenamefont {Harii}, \citenamefont {Oyanagi},\ and\ \citenamefont {Saitoh}}]{Arisawa2019}%
  \BibitemOpen
  \bibfield  {author} {\bibinfo {author} {\bibfnamefont {H.}~\bibnamefont {Arisawa}}, \bibinfo {author} {\bibfnamefont {S.}~\bibnamefont {Daimon}}, \bibinfo {author} {\bibfnamefont {Y.}~\bibnamefont {Oikawa}}, \bibinfo {author} {\bibfnamefont {Y.~J.}\ \bibnamefont {Seo}}, \bibinfo {author} {\bibfnamefont {K.}~\bibnamefont {Harii}}, \bibinfo {author} {\bibfnamefont {K.}~\bibnamefont {Oyanagi}},\ and\ \bibinfo {author} {\bibfnamefont {E.}~\bibnamefont {Saitoh}},\ }\href@noop {} {\bibfield  {journal} {\bibinfo  {journal} {Appl. Phys. Lett.}\ }\textbf {\bibinfo {volume} {114}},\ \bibinfo {pages} {122402} (\bibinfo {year} {2019})}\BibitemShut {NoStop}%
\bibitem [{\citenamefont {Zhu}\ \emph {et~al.}(2017)\citenamefont {Zhu}, \citenamefont {Chang}, \citenamefont {Franson}, \citenamefont {Liu}, \citenamefont {Zhang}, \citenamefont {Johnston-Halperin}, \citenamefont {Wu},\ and\ \citenamefont {Tang}}]{Zhu2017}%
  \BibitemOpen
  \bibfield  {author} {\bibinfo {author} {\bibfnamefont {N.}~\bibnamefont {Zhu}}, \bibinfo {author} {\bibfnamefont {H.}~\bibnamefont {Chang}}, \bibinfo {author} {\bibfnamefont {A.}~\bibnamefont {Franson}}, \bibinfo {author} {\bibfnamefont {T.}~\bibnamefont {Liu}}, \bibinfo {author} {\bibfnamefont {X.}~\bibnamefont {Zhang}}, \bibinfo {author} {\bibfnamefont {E.}~\bibnamefont {Johnston-Halperin}}, \bibinfo {author} {\bibfnamefont {M.}~\bibnamefont {Wu}},\ and\ \bibinfo {author} {\bibfnamefont {H.~X.}\ \bibnamefont {Tang}},\ }\href {https://doi.org/10.1063/1.4986474} {\bibfield  {journal} {\bibinfo  {journal} {Appl. Phys. Lett.}\ }\textbf {\bibinfo {volume} {110}},\ \bibinfo {pages} {252401} (\bibinfo {year} {2017})}\BibitemShut {NoStop}%
\bibitem [{\citenamefont {Heyroth}\ \emph {et~al.}(2019)\citenamefont {Heyroth}, \citenamefont {Hauser}, \citenamefont {Trempler}, \citenamefont {Geyer}, \citenamefont {Syrowatka}, \citenamefont {Dreyer}, \citenamefont {Ebbinghaus}, \citenamefont {Woltersdorf},\ and\ \citenamefont {Schmidt}}]{Heyroth2019}%
  \BibitemOpen
  \bibfield  {author} {\bibinfo {author} {\bibfnamefont {F.}~\bibnamefont {Heyroth}}, \bibinfo {author} {\bibfnamefont {C.}~\bibnamefont {Hauser}}, \bibinfo {author} {\bibfnamefont {P.}~\bibnamefont {Trempler}}, \bibinfo {author} {\bibfnamefont {P.}~\bibnamefont {Geyer}}, \bibinfo {author} {\bibfnamefont {F.}~\bibnamefont {Syrowatka}}, \bibinfo {author} {\bibfnamefont {R.}~\bibnamefont {Dreyer}}, \bibinfo {author} {\bibfnamefont {S.~G.}\ \bibnamefont {Ebbinghaus}}, \bibinfo {author} {\bibfnamefont {G.}~\bibnamefont {Woltersdorf}},\ and\ \bibinfo {author} {\bibfnamefont {G.}~\bibnamefont {Schmidt}},\ }\href {https://doi.org/10.1103/PhysRevApplied.12.054031} {\bibfield  {journal} {\bibinfo  {journal} {Phys. Rev. Appl.}\ }\textbf {\bibinfo {volume} {12}},\ \bibinfo {pages} {054031} (\bibinfo {year} {2019})}\BibitemShut {NoStop}%
\bibitem [{\citenamefont {Taniguchi}\ \emph {et~al.}(2024)\citenamefont {Taniguchi}, \citenamefont {Kitai}, \citenamefont {Yambe}, \citenamefont {Gao}, \citenamefont {Iwamoto},\ and\ \citenamefont {Ota}}]{Taniguchi2024}%
  \BibitemOpen
  \bibfield  {author} {\bibinfo {author} {\bibfnamefont {K.}~\bibnamefont {Taniguchi}}, \bibinfo {author} {\bibfnamefont {T.}~\bibnamefont {Kitai}}, \bibinfo {author} {\bibfnamefont {T.}~\bibnamefont {Yambe}}, \bibinfo {author} {\bibfnamefont {S.}~\bibnamefont {Gao}}, \bibinfo {author} {\bibfnamefont {S.}~\bibnamefont {Iwamoto}},\ and\ \bibinfo {author} {\bibfnamefont {Y.}~\bibnamefont {Ota}},\ }in\ \href@noop {} {\emph {\bibinfo {booktitle} {2024 Conference on Lasers and Electro-Optics Pacific Rim (CLEO-PR)}}}\ (\bibinfo {year} {2024})\BibitemShut {NoStop}%
\bibitem [{\citenamefont {Inoue}\ \emph {et~al.}(2006)\citenamefont {Inoue}, \citenamefont {Fujikawa}, \citenamefont {Baryshev}, \citenamefont {Khanikaev}, \citenamefont {Lim}, \citenamefont {Uchida}, \citenamefont {Aktsipetrov}, \citenamefont {Fedyanin}, \citenamefont {Murzina},\ and\ \citenamefont {Granovsky}}]{Inoue2006}%
  \BibitemOpen
  \bibfield  {author} {\bibinfo {author} {\bibfnamefont {M.}~\bibnamefont {Inoue}}, \bibinfo {author} {\bibfnamefont {R.}~\bibnamefont {Fujikawa}}, \bibinfo {author} {\bibfnamefont {A.}~\bibnamefont {Baryshev}}, \bibinfo {author} {\bibfnamefont {A.}~\bibnamefont {Khanikaev}}, \bibinfo {author} {\bibfnamefont {P.~B.}\ \bibnamefont {Lim}}, \bibinfo {author} {\bibfnamefont {H.}~\bibnamefont {Uchida}}, \bibinfo {author} {\bibfnamefont {O.}~\bibnamefont {Aktsipetrov}}, \bibinfo {author} {\bibfnamefont {A.}~\bibnamefont {Fedyanin}}, \bibinfo {author} {\bibfnamefont {T.}~\bibnamefont {Murzina}},\ and\ \bibinfo {author} {\bibfnamefont {A.}~\bibnamefont {Granovsky}},\ }\href {https://doi.org/10.1088/0022-3727/39/8/R01} {\bibfield  {journal} {\bibinfo  {journal} {J. Phys. D: Appl. Phys.}\ }\textbf {\bibinfo {volume} {39}},\ \bibinfo {pages} {R151} (\bibinfo {year} {2006})}\BibitemShut {NoStop}%
\bibitem [{\citenamefont {Fakhrul}\ \emph {et~al.}(2019)\citenamefont {Fakhrul}, \citenamefont {Tazlaru}, \citenamefont {Beran}, \citenamefont {Zhang}, \citenamefont {Veis},\ and\ \citenamefont {Ross}}]{Fakhrul2019}%
  \BibitemOpen
  \bibfield  {author} {\bibinfo {author} {\bibfnamefont {T.}~\bibnamefont {Fakhrul}}, \bibinfo {author} {\bibfnamefont {S.}~\bibnamefont {Tazlaru}}, \bibinfo {author} {\bibfnamefont {L.}~\bibnamefont {Beran}}, \bibinfo {author} {\bibfnamefont {Y.}~\bibnamefont {Zhang}}, \bibinfo {author} {\bibfnamefont {M.}~\bibnamefont {Veis}},\ and\ \bibinfo {author} {\bibfnamefont {C.~A.}\ \bibnamefont {Ross}},\ }\href {https://doi.org/10.1002/adom.201900056} {\bibfield  {journal} {\bibinfo  {journal} {Adv. Opt. Mater.}\ }\textbf {\bibinfo {volume} {7}},\ \bibinfo {pages} {1900056} (\bibinfo {year} {2019})}\BibitemShut {NoStop}%
\bibitem [{\citenamefont {Inoue}\ \emph {et~al.}(2013)\citenamefont {Inoue}, \citenamefont {Levy},\ and\ \citenamefont {Baryshev}}]{Inoue2013}%
  \BibitemOpen
  \bibfield  {author} {\bibinfo {author} {\bibfnamefont {M.}~\bibnamefont {Inoue}}, \bibinfo {author} {\bibfnamefont {M.}~\bibnamefont {Levy}},\ and\ \bibinfo {author} {\bibfnamefont {A.~V.}\ \bibnamefont {Baryshev}},\ }\href@noop {} {\emph {\bibinfo {title} {Magnetophotonics: From Theory to Applications}}},\ Vol.\ \bibinfo {volume} {190}\ (\bibinfo  {publisher} {Springer},\ \bibinfo {year} {2013})\BibitemShut {NoStop}%
\bibitem [{\citenamefont {Chakravarty}\ \emph {et~al.}(2011)\citenamefont {Chakravarty}, \citenamefont {Levy}, \citenamefont {Jalali}, \citenamefont {Wu},\ and\ \citenamefont {Merzlikin}}]{Chakravarty2011_YIGPHC}%
  \BibitemOpen
  \bibfield  {author} {\bibinfo {author} {\bibfnamefont {A.}~\bibnamefont {Chakravarty}}, \bibinfo {author} {\bibfnamefont {M.}~\bibnamefont {Levy}}, \bibinfo {author} {\bibfnamefont {A.~A.}\ \bibnamefont {Jalali}}, \bibinfo {author} {\bibfnamefont {Z.}~\bibnamefont {Wu}},\ and\ \bibinfo {author} {\bibfnamefont {A.~M.}\ \bibnamefont {Merzlikin}},\ }\href {https://doi.org/10.1103/PhysRevB.84.094202} {\bibfield  {journal} {\bibinfo  {journal} {Phys. Rev. B}\ }\textbf {\bibinfo {volume} {84}},\ \bibinfo {pages} {094202} (\bibinfo {year} {2011})}\BibitemShut {NoStop}%
\bibitem [{\citenamefont {Babinec}\ \emph {et~al.}(2011)\citenamefont {Babinec}, \citenamefont {Choy}, \citenamefont {Smith}, \citenamefont {Khan},\ and\ \citenamefont {Lončar}}]{Babinec2011}%
  \BibitemOpen
  \bibfield  {author} {\bibinfo {author} {\bibfnamefont {T.~M.}\ \bibnamefont {Babinec}}, \bibinfo {author} {\bibfnamefont {J.~T.}\ \bibnamefont {Choy}}, \bibinfo {author} {\bibfnamefont {K.~J.~M.}\ \bibnamefont {Smith}}, \bibinfo {author} {\bibfnamefont {M.}~\bibnamefont {Khan}},\ and\ \bibinfo {author} {\bibfnamefont {M.}~\bibnamefont {Lončar}},\ }\href {https://doi.org/10.1116/1.3520638} {\bibfield  {journal} {\bibinfo  {journal} {J. Vac. Sci. Technol.}\ }\textbf {\bibinfo {volume} {29}},\ \bibinfo {pages} {010601} (\bibinfo {year} {2011})}\BibitemShut {NoStop}%
\bibitem [{COM(2023)}]{COMSOL}%
  \BibitemOpen
  \href@noop {} {\emph {\bibinfo {title} {COMSOL Multiphysics® v. 6.0.}}},\ \bibinfo {organization} {COMSOL AB},\ \bibinfo {address} {Stockholm, Sweden} (\bibinfo {year} {2023})\BibitemShut {NoStop}%
\bibitem [{\citenamefont {Kim}\ \emph {et~al.}(2012)\citenamefont {Kim}, \citenamefont {Ahn},\ and\ \citenamefont {Jang}}]{Kim2012}%
  \BibitemOpen
  \bibfield  {author} {\bibinfo {author} {\bibfnamefont {C.~S.}\ \bibnamefont {Kim}}, \bibinfo {author} {\bibfnamefont {S.~H.}\ \bibnamefont {Ahn}},\ and\ \bibinfo {author} {\bibfnamefont {D.~Y.}\ \bibnamefont {Jang}},\ }\href {https://doi.org/10.1016/j.vacuum.2011.11.004} {\bibfield  {journal} {\bibinfo  {journal} {Vacuum}\ }\textbf {\bibinfo {volume} {86}},\ \bibinfo {pages} {1014} (\bibinfo {year} {2012})}\BibitemShut {NoStop}%
\bibitem [{\citenamefont {Lee}\ \emph {et~al.}(2024)\citenamefont {Lee}, \citenamefont {Yun}, \citenamefont {Lee}, \citenamefont {Lee}, \citenamefont {Lee},\ and\ \citenamefont {Seo}}]{lee2024highqv}%
  \BibitemOpen
  \bibfield  {author} {\bibinfo {author} {\bibfnamefont {J.}~\bibnamefont {Lee}}, \bibinfo {author} {\bibfnamefont {S.}~\bibnamefont {Yun}}, \bibinfo {author} {\bibfnamefont {H.}~\bibnamefont {Lee}}, \bibinfo {author} {\bibfnamefont {C.-H.}\ \bibnamefont {Lee}}, \bibinfo {author} {\bibfnamefont {H.-S.}\ \bibnamefont {Lee}},\ and\ \bibinfo {author} {\bibfnamefont {M.-K.}\ \bibnamefont {Seo}},\ }in\ \href@noop {} {\emph {\bibinfo {booktitle} {CLEO-PR}}},\ \bibinfo {series and number} {Technical Digest}\ (\bibinfo  {publisher} {OPG},\ \bibinfo {year} {2024})\ p.\ \bibinfo {pages} {paper P3\_021}\BibitemShut {NoStop}%
\bibitem [{\citenamefont {Khizroev}\ and\ \citenamefont {Litvinov}(2004)}]{Khizroev2004}%
  \BibitemOpen
  \bibfield  {author} {\bibinfo {author} {\bibfnamefont {S.}~\bibnamefont {Khizroev}}\ and\ \bibinfo {author} {\bibfnamefont {D.}~\bibnamefont {Litvinov}},\ }\href {https://doi.org/10.1088/0957-4484/15/3/R01} {\bibfield  {journal} {\bibinfo  {journal} {Nanotechnology}\ }\textbf {\bibinfo {volume} {15}},\ \bibinfo {pages} {R7} (\bibinfo {year} {2004})}\BibitemShut {NoStop}%
\bibitem [{\citenamefont {Fraser}(2010)}]{Fraser2010}%
  \BibitemOpen
  \bibfield  {author} {\bibinfo {author} {\bibfnamefont {A.~E.}\ \bibnamefont {Fraser}},\ }\emph {\bibinfo {title} {Focused Ion Beam Milled Magnetic Cantilevers}},\ \href@noop {} {\bibinfo {type} {Ph.d. thesis}},\ \bibinfo  {school} {University of Alberta}, \bibinfo {address} {Edmonton, Canada} (\bibinfo {year} {2010})\BibitemShut {NoStop}%
\bibitem [{\citenamefont {Losby}\ \emph {et~al.}(2015)\citenamefont {Losby}, \citenamefont {Sani}, \citenamefont {Grandmont}, \citenamefont {Diao}, \citenamefont {Belov}, \citenamefont {Burgess}, \citenamefont {Compton}, \citenamefont {Hiebert}, \citenamefont {Vick}, \citenamefont {Mohammad}, \citenamefont {Salimi}, \citenamefont {Bridges}, \citenamefont {Thomson},\ and\ \citenamefont {Freeman}}]{Losby2015}%
  \BibitemOpen
  \bibfield  {author} {\bibinfo {author} {\bibfnamefont {J.~E.}\ \bibnamefont {Losby}}, \bibinfo {author} {\bibfnamefont {F.~F.}\ \bibnamefont {Sani}}, \bibinfo {author} {\bibfnamefont {D.~T.}\ \bibnamefont {Grandmont}}, \bibinfo {author} {\bibfnamefont {Z.}~\bibnamefont {Diao}}, \bibinfo {author} {\bibfnamefont {M.}~\bibnamefont {Belov}}, \bibinfo {author} {\bibfnamefont {J.~A.}\ \bibnamefont {Burgess}}, \bibinfo {author} {\bibfnamefont {S.~R.}\ \bibnamefont {Compton}}, \bibinfo {author} {\bibfnamefont {W.~K.}\ \bibnamefont {Hiebert}}, \bibinfo {author} {\bibfnamefont {D.}~\bibnamefont {Vick}}, \bibinfo {author} {\bibfnamefont {K.}~\bibnamefont {Mohammad}}, \bibinfo {author} {\bibfnamefont {E.}~\bibnamefont {Salimi}}, \bibinfo {author} {\bibfnamefont {G.~E.}\ \bibnamefont {Bridges}}, \bibinfo {author} {\bibfnamefont {D.~J.}\ \bibnamefont {Thomson}},\ and\ \bibinfo {author} {\bibfnamefont {M.~R.}\ \bibnamefont {Freeman}},\ }\href {https://doi.org/10.1126/science.aad2449} {\bibfield  {journal} {\bibinfo
  {journal} {Science}\ }\textbf {\bibinfo {volume} {350}},\ \bibinfo {pages} {798} (\bibinfo {year} {2015})}\BibitemShut {NoStop}%
\bibitem [{\citenamefont {Horridge}\ and\ \citenamefont {Tamm}(1969)}]{Horridge1969}%
  \BibitemOpen
  \bibfield  {author} {\bibinfo {author} {\bibfnamefont {G.~A.}\ \bibnamefont {Horridge}}\ and\ \bibinfo {author} {\bibfnamefont {S.~L.}\ \bibnamefont {Tamm}},\ }\href {https://doi.org/10.1126/science.163.3869.817} {\bibfield  {journal} {\bibinfo  {journal} {Science}\ }\textbf {\bibinfo {volume} {163}},\ \bibinfo {pages} {817} (\bibinfo {year} {1969})}\BibitemShut {NoStop}%
\bibitem [{\citenamefont {Hauer}\ \emph {et~al.}(2014)\citenamefont {Hauer}, \citenamefont {Kim}, \citenamefont {Doolin}, \citenamefont {MacDonald}, \citenamefont {Ramp},\ and\ \citenamefont {Davis}}]{Hauer2014}%
  \BibitemOpen
  \bibfield  {author} {\bibinfo {author} {\bibfnamefont {B.~D.}\ \bibnamefont {Hauer}}, \bibinfo {author} {\bibfnamefont {P.~H.}\ \bibnamefont {Kim}}, \bibinfo {author} {\bibfnamefont {C.}~\bibnamefont {Doolin}}, \bibinfo {author} {\bibfnamefont {A.~J.}\ \bibnamefont {MacDonald}}, \bibinfo {author} {\bibfnamefont {H.}~\bibnamefont {Ramp}},\ and\ \bibinfo {author} {\bibfnamefont {J.~P.}\ \bibnamefont {Davis}},\ }\href {https://doi.org/10.1140/epjti4} {\bibfield  {journal} {\bibinfo  {journal} {EPJ tech. instrum.}\ }\textbf {\bibinfo {volume} {1}},\ \bibinfo {pages} {4} (\bibinfo {year} {2014})}\BibitemShut {NoStop}%
\bibitem [{\citenamefont {Kerdoncuff}\ \emph {et~al.}(2015)\citenamefont {Kerdoncuff}, \citenamefont {Hoff}, \citenamefont {Harris}, \citenamefont {Bowen},\ and\ \citenamefont {Andersen}}]{Kerdoncuff2015}%
  \BibitemOpen
  \bibfield  {author} {\bibinfo {author} {\bibfnamefont {H.}~\bibnamefont {Kerdoncuff}}, \bibinfo {author} {\bibfnamefont {U.~B.}\ \bibnamefont {Hoff}}, \bibinfo {author} {\bibfnamefont {G.~I.}\ \bibnamefont {Harris}}, \bibinfo {author} {\bibfnamefont {W.~P.}\ \bibnamefont {Bowen}},\ and\ \bibinfo {author} {\bibfnamefont {U.~L.}\ \bibnamefont {Andersen}},\ }\href {https://doi.org/10.1002/andp.201400171} {\bibfield  {journal} {\bibinfo  {journal} {Ann. Phys.}\ }\textbf {\bibinfo {volume} {527}},\ \bibinfo {pages} {107} (\bibinfo {year} {2015})}\BibitemShut {NoStop}%
\bibitem [{\citenamefont {Xiao}\ \emph {et~al.}(2013)\citenamefont {Xiao}, \citenamefont {Fang}, \citenamefont {Xu}, \citenamefont {Wu},\ and\ \citenamefont {Shen}}]{Xiao2013}%
  \BibitemOpen
  \bibfield  {author} {\bibinfo {author} {\bibfnamefont {Y.~J.}\ \bibnamefont {Xiao}}, \bibinfo {author} {\bibfnamefont {F.~Z.}\ \bibnamefont {Fang}}, \bibinfo {author} {\bibfnamefont {Z.~W.}\ \bibnamefont {Xu}}, \bibinfo {author} {\bibfnamefont {W.}~\bibnamefont {Wu}},\ and\ \bibinfo {author} {\bibfnamefont {X.~C.}\ \bibnamefont {Shen}},\ }\href {https://doi.org/10.1016/j.nimb.2012.12.112} {\bibfield  {journal} {\bibinfo  {journal} {Nucl. Instrum. Methods Phys. Res., Sect. B}\ }\textbf {\bibinfo {volume} {307}},\ \bibinfo {pages} {253} (\bibinfo {year} {2013})}\BibitemShut {NoStop}%
\bibitem [{\citenamefont {Huang}\ \emph {et~al.}(2015)\citenamefont {Huang}, \citenamefont {Loeffler}, \citenamefont {Moeller},\ and\ \citenamefont {Zschech}}]{huang2015tridyn}%
  \BibitemOpen
  \bibfield  {author} {\bibinfo {author} {\bibfnamefont {J.}~\bibnamefont {Huang}}, \bibinfo {author} {\bibfnamefont {M.}~\bibnamefont {Loeffler}}, \bibinfo {author} {\bibfnamefont {W.}~\bibnamefont {Moeller}},\ and\ \bibinfo {author} {\bibfnamefont {E.}~\bibnamefont {Zschech}},\ }\href@noop {} {\bibinfo {title} {Study of ga ion induced amorphization in si during fib using tridyn simulation}},\ \bibinfo {howpublished} {2015 FCMN} (\bibinfo {year} {2015})\BibitemShut {NoStop}%
\bibitem [{\citenamefont {Hrnčíř}\ \emph {et~al.}(2015)\citenamefont {Hrnčíř}, \citenamefont {Obona}, \citenamefont {Petrenec}, \citenamefont {Michalička},\ and\ \citenamefont {Lang}}]{Hrn2015}%
  \BibitemOpen
  \bibfield  {author} {\bibinfo {author} {\bibfnamefont {T.}~\bibnamefont {Hrnčíř}}, \bibinfo {author} {\bibfnamefont {J.~V.}\ \bibnamefont {Obona}}, \bibinfo {author} {\bibfnamefont {M.}~\bibnamefont {Petrenec}}, \bibinfo {author} {\bibfnamefont {J.}~\bibnamefont {Michalička}},\ and\ \bibinfo {author} {\bibfnamefont {C.}~\bibnamefont {Lang}},\ }in\ \href {https://doi.org/10.31399/asm.cp.istfa2015p0065} {\emph {\bibinfo {booktitle} {ISTFA Proceedings}}}\ (\bibinfo {year} {2015})\BibitemShut {NoStop}%
\bibitem [{\citenamefont {Vitale}\ and\ \citenamefont {Sugar}(2022)}]{Vitale2022}%
  \BibitemOpen
  \bibfield  {author} {\bibinfo {author} {\bibfnamefont {S.~M.}\ \bibnamefont {Vitale}}\ and\ \bibinfo {author} {\bibfnamefont {J.~D.}\ \bibnamefont {Sugar}},\ }\href {https://doi.org/10.1017/S1431927622000344} {\bibfield  {journal} {\bibinfo  {journal} {Microsc. Microanal.}\ }\textbf {\bibinfo {volume} {28}},\ \bibinfo {pages} {646} (\bibinfo {year} {2022})}\BibitemShut {NoStop}%
\bibitem [{\citenamefont {Asano}\ \emph {et~al.}(2017)\citenamefont {Asano}, \citenamefont {Ochi}, \citenamefont {Takahashi}, \citenamefont {Kishimoto},\ and\ \citenamefont {Noda}}]{Asano2017}%
  \BibitemOpen
  \bibfield  {author} {\bibinfo {author} {\bibfnamefont {T.}~\bibnamefont {Asano}}, \bibinfo {author} {\bibfnamefont {Y.}~\bibnamefont {Ochi}}, \bibinfo {author} {\bibfnamefont {Y.}~\bibnamefont {Takahashi}}, \bibinfo {author} {\bibfnamefont {K.}~\bibnamefont {Kishimoto}},\ and\ \bibinfo {author} {\bibfnamefont {S.}~\bibnamefont {Noda}},\ }\href {https://doi.org/10.1364/oe.25.001769} {\bibfield  {journal} {\bibinfo  {journal} {Opt. Express}\ }\textbf {\bibinfo {volume} {25}},\ \bibinfo {pages} {1769} (\bibinfo {year} {2017})}\BibitemShut {NoStop}%
\bibitem [{\citenamefont {Yamaguchi}\ \emph {et~al.}(2015)\citenamefont {Yamaguchi}, \citenamefont {Jeon}, \citenamefont {Song}, \citenamefont {Tanaka}, \citenamefont {Asano},\ and\ \citenamefont {Noda}}]{Yamaguchi2015}%
  \BibitemOpen
  \bibfield  {author} {\bibinfo {author} {\bibfnamefont {Y.}~\bibnamefont {Yamaguchi}}, \bibinfo {author} {\bibfnamefont {S.-W.}\ \bibnamefont {Jeon}}, \bibinfo {author} {\bibfnamefont {B.-S.}\ \bibnamefont {Song}}, \bibinfo {author} {\bibfnamefont {Y.}~\bibnamefont {Tanaka}}, \bibinfo {author} {\bibfnamefont {T.}~\bibnamefont {Asano}},\ and\ \bibinfo {author} {\bibfnamefont {S.}~\bibnamefont {Noda}},\ }\href {https://doi.org/10.1364/josab.32.001792} {\bibfield  {journal} {\bibinfo  {journal} {J. Opt. Soc. Am. B}\ }\textbf {\bibinfo {volume} {32}},\ \bibinfo {pages} {1792} (\bibinfo {year} {2015})}\BibitemShut {NoStop}%
\bibitem [{\citenamefont {Zhong}\ \emph {et~al.}(2016)\citenamefont {Zhong}, \citenamefont {Rochman}, \citenamefont {Kindem}, \citenamefont {Miyazono},\ and\ \citenamefont {Faraon}}]{Zhong2016}%
  \BibitemOpen
  \bibfield  {author} {\bibinfo {author} {\bibfnamefont {T.}~\bibnamefont {Zhong}}, \bibinfo {author} {\bibfnamefont {J.}~\bibnamefont {Rochman}}, \bibinfo {author} {\bibfnamefont {J.~M.}\ \bibnamefont {Kindem}}, \bibinfo {author} {\bibfnamefont {E.}~\bibnamefont {Miyazono}},\ and\ \bibinfo {author} {\bibfnamefont {A.}~\bibnamefont {Faraon}},\ }\href {https://doi.org/10.1364/oe.24.000536} {\bibfield  {journal} {\bibinfo  {journal} {Opt. Express}\ }\textbf {\bibinfo {volume} {24}},\ \bibinfo {pages} {536} (\bibinfo {year} {2016})}\BibitemShut {NoStop}%
\bibitem [{\citenamefont {Zhong}\ \emph {et~al.}(2015)\citenamefont {Zhong}, \citenamefont {Kindem}, \citenamefont {Miyazono},\ and\ \citenamefont {Faraon}}]{Zhong2015}%
  \BibitemOpen
  \bibfield  {author} {\bibinfo {author} {\bibfnamefont {T.}~\bibnamefont {Zhong}}, \bibinfo {author} {\bibfnamefont {J.~M.}\ \bibnamefont {Kindem}}, \bibinfo {author} {\bibfnamefont {E.}~\bibnamefont {Miyazono}},\ and\ \bibinfo {author} {\bibfnamefont {A.}~\bibnamefont {Faraon}},\ }\href {https://doi.org/10.1038/ncomms9206} {\bibfield  {journal} {\bibinfo  {journal} {Nat. Commun.}\ }\textbf {\bibinfo {volume} {6}},\ \bibinfo {pages} {8206} (\bibinfo {year} {2015})}\BibitemShut {NoStop}%
\bibitem [{\citenamefont {Zhong}\ \emph {et~al.}(2017)\citenamefont {Zhong}, \citenamefont {Kindem}, \citenamefont {Rochman},\ and\ \citenamefont {Faraon}}]{Zhong2017}%
  \BibitemOpen
  \bibfield  {author} {\bibinfo {author} {\bibfnamefont {T.}~\bibnamefont {Zhong}}, \bibinfo {author} {\bibfnamefont {J.~M.}\ \bibnamefont {Kindem}}, \bibinfo {author} {\bibfnamefont {J.}~\bibnamefont {Rochman}},\ and\ \bibinfo {author} {\bibfnamefont {A.}~\bibnamefont {Faraon}},\ }\href {https://doi.org/10.1038/ncomms14107} {\bibfield  {journal} {\bibinfo  {journal} {Nat. Commun.}\ }\textbf {\bibinfo {volume} {8}},\ \bibinfo {pages} {14107} (\bibinfo {year} {2017})}\BibitemShut {NoStop}%
\bibitem [{\citenamefont {Chan}\ \emph {et~al.}(2012)\citenamefont {Chan}, \citenamefont {Safavi-Naeini}, \citenamefont {Hill}, \citenamefont {Meenehan},\ and\ \citenamefont {Painter}}]{Chan2012}%
  \BibitemOpen
  \bibfield  {author} {\bibinfo {author} {\bibfnamefont {J.}~\bibnamefont {Chan}}, \bibinfo {author} {\bibfnamefont {A.~H.}\ \bibnamefont {Safavi-Naeini}}, \bibinfo {author} {\bibfnamefont {J.~T.}\ \bibnamefont {Hill}}, \bibinfo {author} {\bibfnamefont {S.}~\bibnamefont {Meenehan}},\ and\ \bibinfo {author} {\bibfnamefont {O.}~\bibnamefont {Painter}},\ }\href {https://doi.org/10.1063/1.4747726} {\bibfield  {journal} {\bibinfo  {journal} {Appl. Phys. Lett.}\ }\textbf {\bibinfo {volume} {101}},\ \bibinfo {pages} {081115} (\bibinfo {year} {2012})}\BibitemShut {NoStop}%
\bibitem [{\citenamefont {Dixon}\ and\ \citenamefont {Matthews}(1967)}]{Dixon1967}%
  \BibitemOpen
  \bibfield  {author} {\bibinfo {author} {\bibfnamefont {R.~W.}\ \bibnamefont {Dixon}}\ and\ \bibinfo {author} {\bibfnamefont {H.}~\bibnamefont {Matthews}},\ }\href {https://doi.org/10.1063/1.1754907} {\bibfield  {journal} {\bibinfo  {journal} {Appl. Phys. Lett.}\ }\textbf {\bibinfo {volume} {10}},\ \bibinfo {pages} {195} (\bibinfo {year} {1967})}\BibitemShut {NoStop}%
\bibitem [{\citenamefont {Lynch}\ \emph {et~al.}(1973)\citenamefont {Lynch}, \citenamefont {Dillon},\ and\ \citenamefont {Uitert}}]{Lynch1973}%
  \BibitemOpen
  \bibfield  {author} {\bibinfo {author} {\bibfnamefont {R.~T.}\ \bibnamefont {Lynch}}, \bibinfo {author} {\bibfnamefont {J.~F.}\ \bibnamefont {Dillon}},\ and\ \bibinfo {author} {\bibfnamefont {L.~G.~V.}\ \bibnamefont {Uitert}},\ }\href {https://doi.org/10.1063/1.1661866} {\bibfield  {journal} {\bibinfo  {journal} {J. Appl. Phys.}\ }\textbf {\bibinfo {volume} {44}},\ \bibinfo {pages} {225} (\bibinfo {year} {1973})}\BibitemShut {NoStop}%
\bibitem [{\citenamefont {Li}\ \emph {et~al.}(2015)\citenamefont {Li}, \citenamefont {Cui}, \citenamefont {Feng}, \citenamefont {Huang}, \citenamefont {Huang}, \citenamefont {Liu},\ and\ \citenamefont {Zhang}}]{Li2015}%
  \BibitemOpen
  \bibfield  {author} {\bibinfo {author} {\bibfnamefont {Y.}~\bibnamefont {Li}}, \bibinfo {author} {\bibfnamefont {K.}~\bibnamefont {Cui}}, \bibinfo {author} {\bibfnamefont {X.}~\bibnamefont {Feng}}, \bibinfo {author} {\bibfnamefont {Y.}~\bibnamefont {Huang}}, \bibinfo {author} {\bibfnamefont {Z.}~\bibnamefont {Huang}}, \bibinfo {author} {\bibfnamefont {F.}~\bibnamefont {Liu}},\ and\ \bibinfo {author} {\bibfnamefont {W.}~\bibnamefont {Zhang}},\ }\href {https://doi.org/10.1088/2040-8978/17/4/045001} {\bibfield  {journal} {\bibinfo  {journal} {J. Opt.}\ }\textbf {\bibinfo {volume} {17}},\ \bibinfo {pages} {045001} (\bibinfo {year} {2015})}\BibitemShut {NoStop}%
\bibitem [{\citenamefont {Eichenfield}\ \emph {et~al.}(2009{\natexlab{b}})\citenamefont {Eichenfield}, \citenamefont {Camacho}, \citenamefont {Chan}, \citenamefont {Vahala},\ and\ \citenamefont {Painter}}]{Eichenfield2009_1}%
  \BibitemOpen
  \bibfield  {author} {\bibinfo {author} {\bibfnamefont {M.}~\bibnamefont {Eichenfield}}, \bibinfo {author} {\bibfnamefont {R.}~\bibnamefont {Camacho}}, \bibinfo {author} {\bibfnamefont {J.}~\bibnamefont {Chan}}, \bibinfo {author} {\bibfnamefont {K.~J.}\ \bibnamefont {Vahala}},\ and\ \bibinfo {author} {\bibfnamefont {O.}~\bibnamefont {Painter}},\ }\href {https://doi.org/10.1038/nature08061} {\bibfield  {journal} {\bibinfo  {journal} {Nature}\ }\textbf {\bibinfo {volume} {459}},\ \bibinfo {pages} {550} (\bibinfo {year} {2009}{\natexlab{b}})}\BibitemShut {NoStop}%
\bibitem [{\citenamefont {Chan}\ \emph {et~al.}(2009)\citenamefont {Chan}, \citenamefont {Eichenfield}, \citenamefont {Camacho},\ and\ \citenamefont {Painter}}]{Chan2009}%
  \BibitemOpen
  \bibfield  {author} {\bibinfo {author} {\bibfnamefont {J.}~\bibnamefont {Chan}}, \bibinfo {author} {\bibfnamefont {M.}~\bibnamefont {Eichenfield}}, \bibinfo {author} {\bibfnamefont {R.}~\bibnamefont {Camacho}},\ and\ \bibinfo {author} {\bibfnamefont {O.}~\bibnamefont {Painter}},\ }\href@noop {} {\bibfield  {journal} {\bibinfo  {journal} {Opt. Express.}\ }\textbf {\bibinfo {volume} {17}},\ \bibinfo {pages} {3802} (\bibinfo {year} {2009})}\BibitemShut {NoStop}%
\bibitem [{\citenamefont {Leijssen}\ and\ \citenamefont {Verhagen}(2015)}]{Leijssen2015}%
  \BibitemOpen
  \bibfield  {author} {\bibinfo {author} {\bibfnamefont {R.}~\bibnamefont {Leijssen}}\ and\ \bibinfo {author} {\bibfnamefont {E.}~\bibnamefont {Verhagen}},\ }\href {https://doi.org/10.1038/srep15974} {\bibfield  {journal} {\bibinfo  {journal} {Sci. Rep.}\ }\textbf {\bibinfo {volume} {5}},\ \bibinfo {pages} {15974} (\bibinfo {year} {2015})}\BibitemShut {NoStop}%
\bibitem [{\citenamefont {Graf}\ \emph {et~al.}(2021)\citenamefont {Graf}, \citenamefont {Sharma}, \citenamefont {Huebl},\ and\ \citenamefont {Kusminskiy}}]{Graf2021}%
  \BibitemOpen
  \bibfield  {author} {\bibinfo {author} {\bibfnamefont {J.}~\bibnamefont {Graf}}, \bibinfo {author} {\bibfnamefont {S.}~\bibnamefont {Sharma}}, \bibinfo {author} {\bibfnamefont {H.}~\bibnamefont {Huebl}},\ and\ \bibinfo {author} {\bibfnamefont {S.~V.}\ \bibnamefont {Kusminskiy}},\ }\href {https://doi.org/10.1103/PhysRevResearch.3.013277} {\bibfield  {journal} {\bibinfo  {journal} {Phys. Rev. Res.}\ }\textbf {\bibinfo {volume} {3}},\ \bibinfo {pages} {013277} (\bibinfo {year} {2021})}\BibitemShut {NoStop}%
\bibitem [{\citenamefont {Venkat}\ \emph {et~al.}(2013)\citenamefont {Venkat}, \citenamefont {Kumar}, \citenamefont {Franchin}, \citenamefont {Dmytriiev}, \citenamefont {Mruczkiewicz}, \citenamefont {Fangohr}, \citenamefont {Barman}, \citenamefont {Krawczyk},\ and\ \citenamefont {Prabhakar}}]{Venkat2013}%
  \BibitemOpen
  \bibfield  {author} {\bibinfo {author} {\bibfnamefont {G.}~\bibnamefont {Venkat}}, \bibinfo {author} {\bibfnamefont {D.}~\bibnamefont {Kumar}}, \bibinfo {author} {\bibfnamefont {M.}~\bibnamefont {Franchin}}, \bibinfo {author} {\bibfnamefont {O.}~\bibnamefont {Dmytriiev}}, \bibinfo {author} {\bibfnamefont {M.}~\bibnamefont {Mruczkiewicz}}, \bibinfo {author} {\bibfnamefont {H.}~\bibnamefont {Fangohr}}, \bibinfo {author} {\bibfnamefont {A.}~\bibnamefont {Barman}}, \bibinfo {author} {\bibfnamefont {M.}~\bibnamefont {Krawczyk}},\ and\ \bibinfo {author} {\bibfnamefont {A.}~\bibnamefont {Prabhakar}},\ }\href {https://doi.org/10.1109/TMAG.2012.2206820} {\bibfield  {journal} {\bibinfo  {journal} {IEEE Trans. Magn.}\ }\textbf {\bibinfo {volume} {49}},\ \bibinfo {pages} {524} (\bibinfo {year} {2013})}\BibitemShut {NoStop}%
\bibitem [{\citenamefont {Vansteenkiste}\ \emph {et~al.}(2014)\citenamefont {Vansteenkiste}, \citenamefont {Leliaert}, \citenamefont {Dvornik}, \citenamefont {Helsen}, \citenamefont {Garcia-Sanchez},\ and\ \citenamefont {Waeyenberge}}]{Vansteenkiste2014}%
  \BibitemOpen
  \bibfield  {author} {\bibinfo {author} {\bibfnamefont {A.}~\bibnamefont {Vansteenkiste}}, \bibinfo {author} {\bibfnamefont {J.}~\bibnamefont {Leliaert}}, \bibinfo {author} {\bibfnamefont {M.}~\bibnamefont {Dvornik}}, \bibinfo {author} {\bibfnamefont {M.}~\bibnamefont {Helsen}}, \bibinfo {author} {\bibfnamefont {F.}~\bibnamefont {Garcia-Sanchez}},\ and\ \bibinfo {author} {\bibfnamefont {B.~V.}\ \bibnamefont {Waeyenberge}},\ }\href {https://doi.org/10.1063/1.4899186} {\bibfield  {journal} {\bibinfo  {journal} {AIP Adv.}\ }\textbf {\bibinfo {volume} {4}},\ \bibinfo {pages} {107133} (\bibinfo {year} {2014})}\BibitemShut {NoStop}%
\bibitem [{\citenamefont {Pitanti}\ \emph {et~al.}(2015)\citenamefont {Pitanti}, \citenamefont {Fink}, \citenamefont {Safavi-Naeini}, \citenamefont {Hill}, \citenamefont {Lei}, \citenamefont {Tredicucci},\ and\ \citenamefont {Painter}}]{Pitanti2015}%
  \BibitemOpen
  \bibfield  {author} {\bibinfo {author} {\bibfnamefont {A.}~\bibnamefont {Pitanti}}, \bibinfo {author} {\bibfnamefont {J.~M.}\ \bibnamefont {Fink}}, \bibinfo {author} {\bibfnamefont {A.~H.}\ \bibnamefont {Safavi-Naeini}}, \bibinfo {author} {\bibfnamefont {J.~T.}\ \bibnamefont {Hill}}, \bibinfo {author} {\bibfnamefont {C.~U.}\ \bibnamefont {Lei}}, \bibinfo {author} {\bibfnamefont {A.}~\bibnamefont {Tredicucci}},\ and\ \bibinfo {author} {\bibfnamefont {O.}~\bibnamefont {Painter}},\ }\href {https://doi.org/10.1364/oe.23.003196} {\bibfield  {journal} {\bibinfo  {journal} {Opt. Express}\ }\textbf {\bibinfo {volume} {23}},\ \bibinfo {pages} {3196} (\bibinfo {year} {2015})}\BibitemShut {NoStop}%
\bibitem [{\citenamefont {Fink}\ \emph {et~al.}(2016)\citenamefont {Fink}, \citenamefont {Kalaee}, \citenamefont {Pitanti}, \citenamefont {Norte}, \citenamefont {Heinzle}, \citenamefont {Davanço}, \citenamefont {Srinivasan},\ and\ \citenamefont {Painter}}]{Fink2016}%
  \BibitemOpen
  \bibfield  {author} {\bibinfo {author} {\bibfnamefont {J.~M.}\ \bibnamefont {Fink}}, \bibinfo {author} {\bibfnamefont {M.}~\bibnamefont {Kalaee}}, \bibinfo {author} {\bibfnamefont {A.}~\bibnamefont {Pitanti}}, \bibinfo {author} {\bibfnamefont {R.}~\bibnamefont {Norte}}, \bibinfo {author} {\bibfnamefont {L.}~\bibnamefont {Heinzle}}, \bibinfo {author} {\bibfnamefont {M.}~\bibnamefont {Davanço}}, \bibinfo {author} {\bibfnamefont {K.}~\bibnamefont {Srinivasan}},\ and\ \bibinfo {author} {\bibfnamefont {O.}~\bibnamefont {Painter}},\ }\href {https://doi.org/10.1038/ncomms12396} {\bibfield  {journal} {\bibinfo  {journal} {Nat. Commun.}\ }\textbf {\bibinfo {volume} {7}},\ \bibinfo {pages} {12396} (\bibinfo {year} {2016})}\BibitemShut {NoStop}%
\bibitem [{\citenamefont {Dieterle}\ \emph {et~al.}(2016)\citenamefont {Dieterle}, \citenamefont {Kalaee}, \citenamefont {Fink},\ and\ \citenamefont {Painter}}]{Dieterle2016}%
  \BibitemOpen
  \bibfield  {author} {\bibinfo {author} {\bibfnamefont {P.~B.}\ \bibnamefont {Dieterle}}, \bibinfo {author} {\bibfnamefont {M.}~\bibnamefont {Kalaee}}, \bibinfo {author} {\bibfnamefont {J.~M.}\ \bibnamefont {Fink}},\ and\ \bibinfo {author} {\bibfnamefont {O.}~\bibnamefont {Painter}},\ }\href {https://doi.org/10.1103/PhysRevApplied.6.014013} {\bibfield  {journal} {\bibinfo  {journal} {Phys. Rev. Appl.}\ }\textbf {\bibinfo {volume} {6}},\ \bibinfo {pages} {014013} (\bibinfo {year} {2016})}\BibitemShut {NoStop}%
\bibitem [{\citenamefont {Li}\ \emph {et~al.}(2019)\citenamefont {Li}, \citenamefont {Polakovic}, \citenamefont {Wang}, \citenamefont {Xu}, \citenamefont {Lendinez}, \citenamefont {Zhang}, \citenamefont {Ding}, \citenamefont {Khaire}, \citenamefont {Saglam}, \citenamefont {Divan}, \citenamefont {Pearson}, \citenamefont {Kwok}, \citenamefont {Xiao}, \citenamefont {Novosad}, \citenamefont {Hoffmann},\ and\ \citenamefont {Zhang}}]{Li2019}%
  \BibitemOpen
  \bibfield  {author} {\bibinfo {author} {\bibfnamefont {Y.}~\bibnamefont {Li}}, \bibinfo {author} {\bibfnamefont {T.}~\bibnamefont {Polakovic}}, \bibinfo {author} {\bibfnamefont {Y.~L.}\ \bibnamefont {Wang}}, \bibinfo {author} {\bibfnamefont {J.}~\bibnamefont {Xu}}, \bibinfo {author} {\bibfnamefont {S.}~\bibnamefont {Lendinez}}, \bibinfo {author} {\bibfnamefont {Z.}~\bibnamefont {Zhang}}, \bibinfo {author} {\bibfnamefont {J.}~\bibnamefont {Ding}}, \bibinfo {author} {\bibfnamefont {T.}~\bibnamefont {Khaire}}, \bibinfo {author} {\bibfnamefont {H.}~\bibnamefont {Saglam}}, \bibinfo {author} {\bibfnamefont {R.}~\bibnamefont {Divan}}, \bibinfo {author} {\bibfnamefont {J.}~\bibnamefont {Pearson}}, \bibinfo {author} {\bibfnamefont {W.~K.}\ \bibnamefont {Kwok}}, \bibinfo {author} {\bibfnamefont {Z.}~\bibnamefont {Xiao}}, \bibinfo {author} {\bibfnamefont {V.}~\bibnamefont {Novosad}}, \bibinfo {author} {\bibfnamefont {A.}~\bibnamefont {Hoffmann}},\ and\ \bibinfo {author} {\bibfnamefont {W.}~\bibnamefont {Zhang}},\
  }\href {https://doi.org/10.1103/PhysRevLett.123.107701} {\bibfield  {journal} {\bibinfo  {journal} {PRL.}\ }\textbf {\bibinfo {volume} {123}},\ \bibinfo {pages} {107701} (\bibinfo {year} {2019})}\BibitemShut {NoStop}%
\bibitem [{\citenamefont {Verma}\ \emph {et~al.}(2024)\citenamefont {Verma}, \citenamefont {Maurya}, \citenamefont {Singh},\ and\ \citenamefont {Bhoi}}]{Verma2024}%
  \BibitemOpen
  \bibfield  {author} {\bibinfo {author} {\bibfnamefont {S.}~\bibnamefont {Verma}}, \bibinfo {author} {\bibfnamefont {A.}~\bibnamefont {Maurya}}, \bibinfo {author} {\bibfnamefont {R.}~\bibnamefont {Singh}},\ and\ \bibinfo {author} {\bibfnamefont {B.}~\bibnamefont {Bhoi}},\ }\href {https://doi.org/10.1007/s10948-024-06721-w} {\bibfield  {journal} {\bibinfo  {journal} {J. Supercond. Nov. Magn.}\ }\textbf {\bibinfo {volume} {37}},\ \bibinfo {pages} {1163} (\bibinfo {year} {2024})}\BibitemShut {NoStop}%
\bibitem [{\citenamefont {Bhoi}\ \emph {et~al.}(2017)\citenamefont {Bhoi}, \citenamefont {Kim}, \citenamefont {Kim}, \citenamefont {Cho},\ and\ \citenamefont {Kim}}]{Bhoi2017}%
  \BibitemOpen
  \bibfield  {author} {\bibinfo {author} {\bibfnamefont {B.}~\bibnamefont {Bhoi}}, \bibinfo {author} {\bibfnamefont {B.}~\bibnamefont {Kim}}, \bibinfo {author} {\bibfnamefont {J.}~\bibnamefont {Kim}}, \bibinfo {author} {\bibfnamefont {Y.~J.}\ \bibnamefont {Cho}},\ and\ \bibinfo {author} {\bibfnamefont {S.~K.}\ \bibnamefont {Kim}},\ }\href {https://doi.org/10.1038/s41598-017-12215-8} {\bibfield  {journal} {\bibinfo  {journal} {Sci. Rep.}\ }\textbf {\bibinfo {volume} {7}},\ \bibinfo {pages} {11930} (\bibinfo {year} {2017})}\BibitemShut {NoStop}%
\bibitem [{\citenamefont {Gonzalez-Ballestero}\ \emph {et~al.}(2020)\citenamefont {Gonzalez-Ballestero}, \citenamefont {H\"ummer}, \citenamefont {Gieseler},\ and\ \citenamefont {Romero-Isart}}]{Gonzales2020}%
  \BibitemOpen
  \bibfield  {author} {\bibinfo {author} {\bibfnamefont {C.}~\bibnamefont {Gonzalez-Ballestero}}, \bibinfo {author} {\bibfnamefont {D.}~\bibnamefont {H\"ummer}}, \bibinfo {author} {\bibfnamefont {J.}~\bibnamefont {Gieseler}},\ and\ \bibinfo {author} {\bibfnamefont {O.}~\bibnamefont {Romero-Isart}},\ }\href {https://doi.org/10.1103/PhysRevB.101.125404} {\bibfield  {journal} {\bibinfo  {journal} {Phys. Rev. B}\ }\textbf {\bibinfo {volume} {101}},\ \bibinfo {pages} {125404} (\bibinfo {year} {2020})}\BibitemShut {NoStop}%
\bibitem [{\citenamefont {Zhang}\ \emph {et~al.}(2016{\natexlab{b}})\citenamefont {Zhang}, \citenamefont {Zou}, \citenamefont {Jiang},\ and\ \citenamefont {Tang}}]{Zhang20163dcav}%
  \BibitemOpen
  \bibfield  {author} {\bibinfo {author} {\bibfnamefont {X.}~\bibnamefont {Zhang}}, \bibinfo {author} {\bibfnamefont {C.}~\bibnamefont {Zou}}, \bibinfo {author} {\bibfnamefont {L.}~\bibnamefont {Jiang}},\ and\ \bibinfo {author} {\bibfnamefont {H.~X.}\ \bibnamefont {Tang}},\ }\href {https://doi.org/10.1063/1.4939134} {\bibfield  {journal} {\bibinfo  {journal} {J. Appl. Phys.}\ }\textbf {\bibinfo {volume} {119}},\ \bibinfo {pages} {023905} (\bibinfo {year} {2016}{\natexlab{b}})}\BibitemShut {NoStop}%
\bibitem [{\citenamefont {Kostylev}\ \emph {et~al.}(2016)\citenamefont {Kostylev}, \citenamefont {Goryachev},\ and\ \citenamefont {Tobar}}]{Kostylev2016}%
  \BibitemOpen
  \bibfield  {author} {\bibinfo {author} {\bibfnamefont {N.}~\bibnamefont {Kostylev}}, \bibinfo {author} {\bibfnamefont {M.}~\bibnamefont {Goryachev}},\ and\ \bibinfo {author} {\bibfnamefont {M.~E.}\ \bibnamefont {Tobar}},\ }\href {https://doi.org/10.1063/1.4941730} {\bibfield  {journal} {\bibinfo  {journal} {Appl. Phys. Lett.}\ }\textbf {\bibinfo {volume} {108}},\ \bibinfo {pages} {023905} (\bibinfo {year} {2016})}\BibitemShut {NoStop}%
\end{thebibliography}%
\end{document}